\documentclass[fleqn,usenatbib]{mnras}

\usepackage{mathptmx}
\usepackage[T1]{fontenc}

\DeclareRobustCommand{\VAN}[3]{#2}
\let\VANthebibliography\thebibliography
\def\thebibliography{\DeclareRobustCommand{\VAN}[3]{##3}\VANthebibliography}

\usepackage{graphicx}	% Including figure files
\usepackage{amsmath}	% Advanced maths commands
\usepackage{amssymb}	% Extra maths symbols

\title[The critical properties of backsplash galaxies]{There and back again: understanding the critical properties of backsplash galaxies}
\author[Josh Borrow et al.]{
Josh Borrow,$^{1}$\thanks{E-mail: josh@joshborrow.com, borrowj@mit.edu (JB)}
Mark Vogelsberger,$^{1}$
Stephanie O'Neil,$^{1}$
Michael A. McDonald$^{1}$ and
Aaron Smith$^{1}$\thanks{NHFP Einstein Fellow.}
\\
$^{1}$Department of Physics and Kavli Institute for Astrophysics and Space Research, Massachusetts Institute of Technology, Cambridge, MA 02139, USA\\
}

\date{Accepted XXX. Received YYY; in original form ZZZ}

\pubyear{2022}

\begin{document}
\label{firstpage}
\pagerange{\pageref{firstpage}--\pageref{lastpage}}
\maketitle

% Abstract of the paper
\begin{abstract}
	Backsplash galaxies are galaxies that once resided inside a cluster, and
	have migrated back oustide as they move towards the apocentre of their
	orbit. The kinematic properties of these galaxies are well understood,
	thanks to the significant study of backsplashers in dark matter-only
	simulations, but their intrinsic properties are not well constrained due to
	modelling uncertainties in sub-grid physics, ram pressure stripping,
	dynamical friction, and tidal forces. In this paper, we use the
	IllustrisTNG300-1 simulation, with a baryonic resolution of $M_{\rm b}
	\approx 1.1\times 10^7$ M$_\odot$, to study backsplash galaxies around 1302
	isolated galaxy clusters with mass $10^{13.0} < M_{\rm 200,mean} /  {\rm M}_\odot<
	10^{15.5}$. We employ a decision tree classifier to extract
	features of galaxies that make them likely to be backsplash galaxies,
	compared to nearby field galaxies, and find that backsplash galaxies have
	low gas fractions, high mass-to-light ratios, large stellar sizes, and low
	black hole occupation fractions. We investigate in detail the origins of
	these large sizes, and hypothesise their origins are linked to the tidal
	environments in the cluster. We show that the black hole recentreing scheme
	employed in many cosmological simulations leads to the loss of black holes
	from galaxies accreted into clusters, and suggest improvements to these
	models. Generally, we find that backsplash galaxies are a useful population
	to test and understand numerical galaxy formation models due to their challenging
	environments and evolutionary pathways that interact with poorly constrained physics.
\end{abstract}

% Select between one and six entries from the list of approved keywords.
% Don't make up new ones.

\section{Introduction}

Understanding the assembly and composition of galaxy clusters has been a topic
of research for decades. Within the Lambda-CDM cosmological paradigm, these
massive systems are understood to be hierachically created from many smaller
substructures.  Galaxy clusters are assembled when many subhaloes, containing
galaxies, accrete together to form self-bound strucutres of halo mass $M_{\rm
200, mean} > 10^{13}$ M$_\odot$. Here $M_{\rm 200, mean}$ is defined as the mass
enclosed within $R_{\rm 200, mean}$, the radius enclosing 200 times the mean
density of the Universe at that epoch. What happens when one of these galaxies
leaves the cluster again?  Is such a process possible? If so, what do these
galaxies look like and why?  This paper explores the life and times of such
`backsplash' galaxies, through the use of the TNG300 cosmological galaxy
formation simulation, and attempts to address these open questions.

Backsplash galaxies were first studied in early theoretical work by
\citet{Balogh2000}, using mass accretion histories from N-body simulations
and an analytical model to explain star formation gradients in the outskirts of
galaxy clusters. The general understanding was that galaxy clusters quench
galaxies, and galaxies that fall into clusters have a chance of `bouncing' back
out of the clusters to an apocentre greater than the virial radius $R_{\rm
vir}$, provided they  have enough orbital energy. Hence, in the outskirts of the
clusters, star forming galaxies would be mixed with these quenched backsplash
galaxies, with the number of backsplashers dropping off with radius but the
density of the infalling population remaining relatively constant with radius, giving rise
to a gradient in star formation activity even further out than $R_{\rm vir}$.

This early work was complimented by theoretical and observational studies by
\citet{Mamon2004} and \citet{Sanchis2004} looking at neutral hydrogen (HI)
deficient galaxies near the Virgo cluster. It was not until \citet{Gill2005},
however, that the term `backsplash galaxy' was actually coined, which was the
first systematic study evaluating the expected kinematic (i.e. including
velocity-space information) properties of these interlopers. \citet{Gill2005}
used N-body simulations to predict that 50\% of galaxies in the range $1 < R /
R_{\rm vir} < 2$ would be backsplash galaxies, and that these galaxies had close
pericentre passages to the centre of the cluster before rebounding. Within a
month, \citet{Kilborn2005} provided observational evidence of two HI-poor
galaxies within this radial range around the NGC 1566 group, further suggesting
that such galaxies could indeed exist in the real Universe.

Later theoretical and observational work by \citet{Ludlow2009}, \citet{Hansen2009},
\citet{Knebe2011}, and \citet{Bahe2013}, using a variety of techniques,
cemented the idea that backsplash galaxies are a cause of colour gradients
in galaxies out to multiple times $R_{\rm 200, mean}$. Notably, \citet{Knebe2011}
found that the processing and production of backsplash subhaloes even occurs on
dwarf galaxy scales, in simulations of the local group.

Much of the work investigating the interaction of cluster galaxies and their
hosts has employed the use of semi-analytical models, which use a base dark
matter-only simulation to generate accretion histories, followed by an
analytical model for studying the baryonic component. Simulating galaxy clusters
with a full hydrodynamical treatment is incredibly complex, with authors turning
to zoom simulations (where only a small fraction of the volume is
hydrodynamically active), or sacrificing numerical resolution significantly
\citep{Barnes2017, McCarthy2017, Ragone-Figueroa2018, Haggar2020, Bassini2020}.
Additionally, the processes involved in the evolution and production of
backsplash galaxies, including ram-pressure stripping and dynamical friction,
are still not fully understood \citep{Simpson2018,vandenBosch2018b}. The
subhaloes that are able to rebound from the cluster are typically of lower mass,
with subhaloes with total halo mass $M_{\rm H} / M_{\rm 200, mean} > 0.01$
strongly impacted by dynamical friction leading to rapid merging with the
central galaxy in the cluster \citep{Bakels2021}.

Despite these potential numerical pitfalls, there has recently been renewed
interest in backsplash galaxies as a potential formation mechanism of
ultra-diffuse galaxies \citep[UDGs, see e.g.][]{vanDokkum2015, Koda2015,
Tremmel2020, Jones2021}. UDGs are low surface-brightness dwarf galaxies,
typically with larger stellar sizes than field dwarfs. Many UDGs are located
near clusters and have intrinsic properties similar to those expected of
backsplash galaxies \citep{Benavides2021,Trujillo2021}. Additionally, the
challenges of simulating backsplash galaxies make them an excellent testing
ground for the accuracy of numerical models. Modern galaxy formation models are
currently able to reproduce a statistcally representative sample of simulated
galaxies, but typically struggle to reproduce the observed properties of
`extremophile' galaxies that reside in extreme environments
\citep{Somerville2015,Vogelsberger2020,delosRios2021}.

Recently, \citet{Farid2022} used a random forest algorithm to classify the
membership properties of cluster galaxies, and was able to perform such
classification with high ($>80\%$) accuracy when employing phase-space
information. In practice, however, phase-space information is difficult to
generate as typically only velocities along the line of sight (from redshift
information) can be observed. To reach this accuracy, they additionally augment
their model by using the specific star formation rate (sSFR) of galaxies, but a
notable inaccuracy in their model is that they underpredict the classification
of orbiting galaxies at radii $1.0 < R / R_{\rm 200, mean} <2 .0$ (i.e. their
model underclassifies backsplash galaxies) when using a hard classification
scheme. 

In this paper, we aim to classify galaxies near clusters into two categories,
infalling (never within $R_{\rm 200, mean}$ of a cluster), and backsplash (has
been within $R_{\rm 200, mean}$ of a cluster, but are now further away). We use
the IllustrisTNG suite of cosmological galaxy formation simulations, and employ
the largest volume simulation with a comoving box-size of $(300 {\rm Mpc})^3$.
Through the development of a classification model for the galaxies based upon
their intrinsic properties, we aim to identify which features of backsplash
galaxies make them unique from the general field population, and use these
features to understand their formation and evolution.

The rest of this paper is organised as follows: In \S \ref{sec:sim}, we give a
brief overview of the IllustrisTNG model and simulations. In \S
\ref{sec:selection}, we describe our cluster and galaxy selection process, and
discuss some basic properties of backsplash galaxies in the TNG model. In \S
\ref{sec:classifier}, we use a decision tree machine learning model to
understand which properties of galaxies can be used to classify them as
backsplashers or infalling. In \S \ref{sec:sizes}, we discuss the origins of the
larger sizes of backsplash galaxies, and in \S \ref{sec:blackholes}, we aim to
understand the low supermassive back hole occupation fraction and masses in
backsplash galaxies. Finally, in \S \ref{sec:conclusions}, we give concluding
remarks.
\section{The IllustrisTNG Simulation Suite}
\label{sec:sim}

In this work we use simulations from the IllustrisTNG suite of cosmological
galaxy formation simulations, released in \citet{Nelson2019}. In particular,
we use data from the public release of the IllustrisTNG300-1 simulation
\citep{Pillepich2018b, Springel2018, Nelson2018, Naiman2018, Marinacci2018}.
The simulation is henceforth referred to as TNG300, and follows a 
(300 Mpc)$^3$ comoving, representative, volume of universe from just
after the big bang down to the present day at $z=0$. All results in this
paper should be assumed to be at $z=0$ unless otherwise stated.

The TNG300 volume initially contains $2500^3$ gas cells of target mass
$1.1\times10^7$ M$_\odot$ and $2500^3$ dark matter particles of fixed mass $5.9
\times 10^7$ M$_\odot$. These particles are evolved over cosmic time using the
continuum magnetohydrodynamics (MHD) and gravity code {\textsc{Arepo}}
\citep{Springel2011,Weinberger2020}. {\textsc{Arepo}} is a massively parallel
code that uses a moving Voronoi mesh coupled with a finite volume method to
compute forces and fluxes originating from MHD, along with a combined tree and
particle mesh (TreePM) method to calculate perodic and non-periodic
self-gravitational forces.

Coupled to the base {\textsc{Arepo}} code is the IllustrisTNG galaxy formation
sub-grid model, a sucessor to the Illustris galaxy formation model
\citep{Vogelsberger2013, Vogelsberger2014b}. Galaxy formation
sub-grid models are required to include physics
that occurs below the resolution scale. Within the TNG model, there are
prescriptions for radiative gas cooling \citep{Wiersma2009}, star formation and
stellar feedback \citep{Springel2003}, and (supermassive) black hole formation
and active galactic nuclei feedback \citep{Weinberger2017}. The simulation was
performed assuming a cosmology conforming to the 2015 release from the Planck
satellite, with $\Omega_{\Lambda,0} = 0.6911$, $\Omega_{\rm m, 0} = 0.3089$, 
$\Omega_{\rm b, 0} = 0.0486$, $\sigma_8 = 0.8159$, $n_{\rm s} = 0.9667$, and
$h=0.6774$, with the symbols having their usual meaning
\citep{PlanckCollaboration2016}. More information on the specific implementation
details is available in the data release paper and associated webpages
\citep{Nelson2019}.

Here we primarily use reduced data from the available {\textsc{SubFind}} group
catalouges.  {\textsc{SubFind}} primarily uses a friends-of-friends (FoF)
algorithm \citep{Springel2005, Dolag2009} to idenfify haloes and their
constituent subhaloes. In TNG300, the initial FoF search uses a typical default
linking length of $b=0.2$, before the {\textsc{SubFind}} galaxy finder algorithm
is ran. For each of the 100 TNG300 snapshots, a group catalogue is created
containing properties of galaxies (subhaloes) and groups (haloes). These group
catalogues are used, along with the {\textsc{L-HaloTree}} algorithm
\citep{Springel2005}, to follow the paths of individual structures over time and
build up a merger tree of haloes. 
\begin{figure*}
    \centering
    \includegraphics{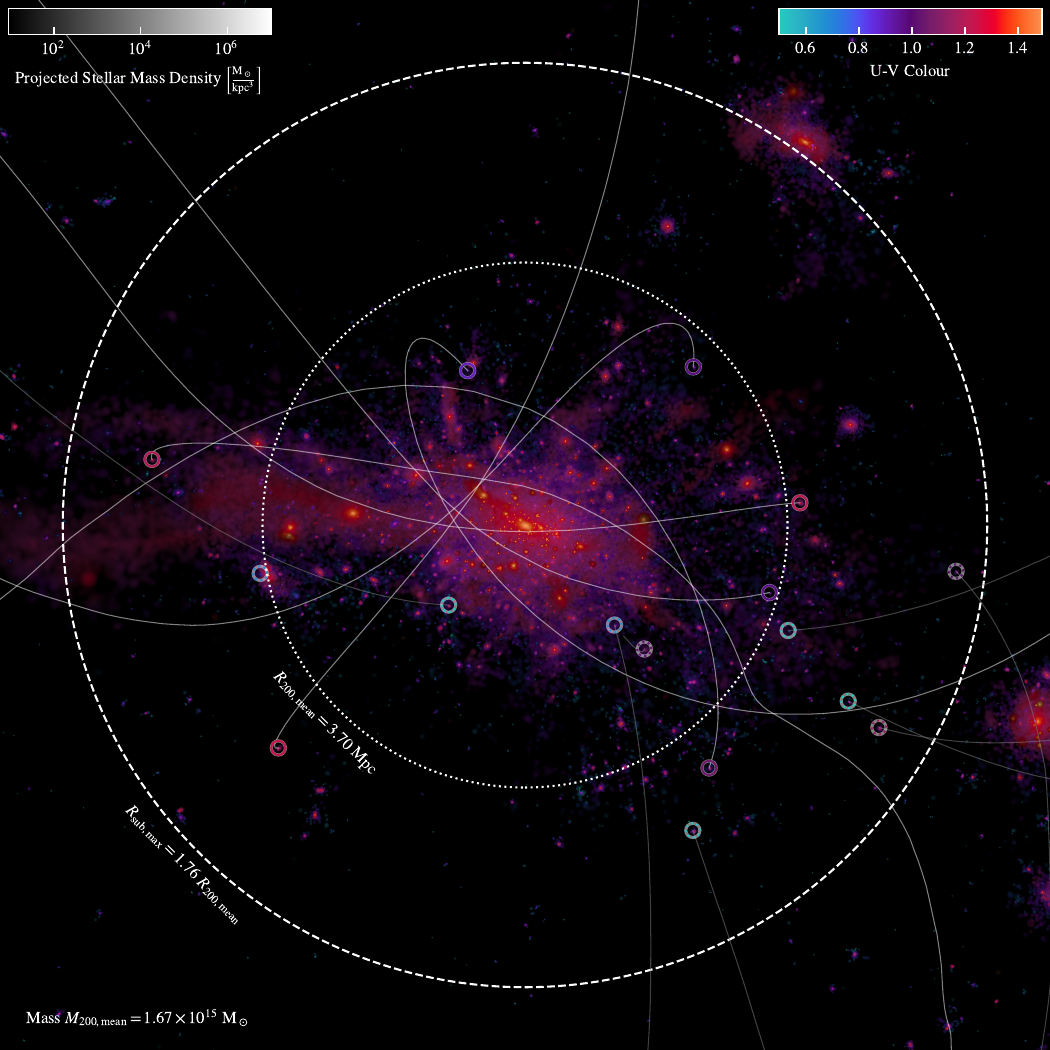}
    \caption{Shows a $10^{15}$ ${\rm M}_\odot$ cluster from the Illustris
    TNG300 simulation. The background image shows the stellar density
    (luminance of the image; i.e. brighter regions have a higher projected
    stellar density), with the colour showing the mean, mass-weighted, $U-V$
    colour along the line of sight. Each ring shows a galaxy that currently
    resides outside the radius $R_{\rm 200, mean}$ (dotted white line), with the
    rings coloured by the $U-V$ colour of the substructure.  If the ring is
    dotted, it represents an infalling galaxy (note that these are typically
    bluer), and if solid it is a backsplash galaxy. The white dashed line shows
    the maximal (3D) radius that a backsplash substructure can be found at for
    this cluster ($R_{\rm sub, max}$). Finally, the thin light lines show the
    paths of the highlighted galaxies.}
    \label{fig:fancyimage}
\end{figure*}

\section{Selecting Backsplash Galaxies}
\label{sec:selection}

As we were motivated to study backsplash galaxies by the study of the
splashback radius of galaxy clusters, we use the same sample of
clusters as in \citet{O'Neil2022}. These are chosen to be isolated
galaxy clusters with a mass $M_{\rm 200, mean} > 10^{13}$ M$_\odot$.
This sample is constructed by ensuring that the clusters are the most
massive within $10 R_{\rm 200, mean}$ of any nearby cluster. This is
ensured by, for every candidate in the $z=0$ snapshot, looping over all
other candidates within $10 R_{\rm 200, mean}$. In this loop, if any
nearby candidate has a mass lower than the cluster being considered, it
is marked as removed from the sample.

The advantage of this approach (as compared to only looping over nearby
members and asserting there must be no halo of greater mass) is that if, say,
there is a $M_{\rm 200, mean} = 10^{13}$ M$_\odot$ galaxy cluster near a $M_{\rm
200, mean} = 10^{15}$ M$_\odot$ cluster, the lower mass one (which will be
significantly impacted by the nearby, more massive, cluster, despite the
$10^{13}$ M$_\odot$ cluster not `seeing' the $10^{15}$ M$_\odot$ cluster in its
radial search) is not included in our sample.

For each isolated cluster we then employ the {\textsc{L-HaloTree}} merger
trees, and track the first progenitor (most massive subhalo) of the cluster
back in time. To ensure that we are able to accurately track the movement of
the cluster back in time (and as such reconstruct the paths of its satellites
accurately), we apply an additional set of constraints:
\begin{itemize}
    \item The first progenitor of the cluster must be traceable through at least
          the most recent 40 snapshots (at least back to $z=0.7$, or approximately
          half a Hubble time). This ensures we have a stable enough cluster to
          follow infalling galaxies for at least a handful of cluster crossing times
          (typically $t_{\rm cross} \approx 1-2$ Gyr).
    \item The cluster must not move more than 10 Mpc per Gyr (comoving) relative to
          the fixed box, to ensure that it corresponds to the local potential maximum.
          This works alongside our isolated galaxy constraint to ensure that as few
          as possible backsplash galaxies are transferred between clusters and experience
          interactions with more than one cluster. In particular, this could skew
          our results since galaxies that have passed through a different cluster
          would not be identified as backsplash despite being subject to a
          cluster environment similar to backsplash galaxies of its current host
          halo.
    \item In a similar fashion, there must be less than a 0.5 Mpc offset between
          where the cluster is predicted to be based upon its velocity and time
          between snapshots relative to its position in the next snapshot.
\end{itemize}
This set of constraints esnures that the clusters move smoothly, and are the
most dominant halo within their local neighborhood, to ensure a clean sample of
infalling and backsplashing galaxies. The final selection of clusters rejects
around 100 compared to \citet{O'Neil2022}, with 1302 clusters remaining.

To track the galaxies resident in, and near, these clusters, we again employ the
{\textsc{L-HaloTree}} merger trees. We identify all galaxies within the range $1
< R_{\rm 200, mean} < 10$ of each cluster at $z=0$, and track their first
progenitor (whether it is bound to, enters, or is assoicated with the merger
tree of the cluster or not) back as far as it can be identified by the tree. We
apply a stellar mass cut of $M_* > 10^8$ M$_\odot$, only tracking substructure
that has a mass higher than this at $z=0$ back in time, to ensure that all
substructure would be visible and to ease computational load, though such a
tracking procedure is not particularly computationally complex.

We then assign galaxies a label of `backsplash' or `infalling' based on whether
they have come within $R_{\rm 200, mean}$ of the cluster (at the relevant
redshift) at any time. In Fig. \ref{fig:fancyimage}, we show the tracks of some
example backsplash and infalling galaxies, alongside a massive galaxy cluster
($M_{\rm 200, mean} = 1.67 \times 10^{15}$ M$_\odot$). The infalling galaxies,
indicated by dotted circles, have yet to interact with the cluster (note that this
is a projection along the $z$-axis of the simulation), whereas the backsplash
galaxies (indicated by solid rings) have had a close periapsis with the cluster.
Notable in this image is the dense prescence of galaxies on the opposing side of
the cluster to the infalling filament (on the left of the image). We also
identify the maximal radius at which any backsplashing substructure is found at,
$R_{\rm sub, max}$.

From Fig. \ref{fig:fancyimage} it is clear that we expect some divergence in the
properties of backsplash and infalling galaxies. It is already clear that the
infalling galaxies are much bluer, with backsplash galaxies being red (i.e. a
high $U-V$ colour).

\begin{figure}
    \centering
    \includegraphics{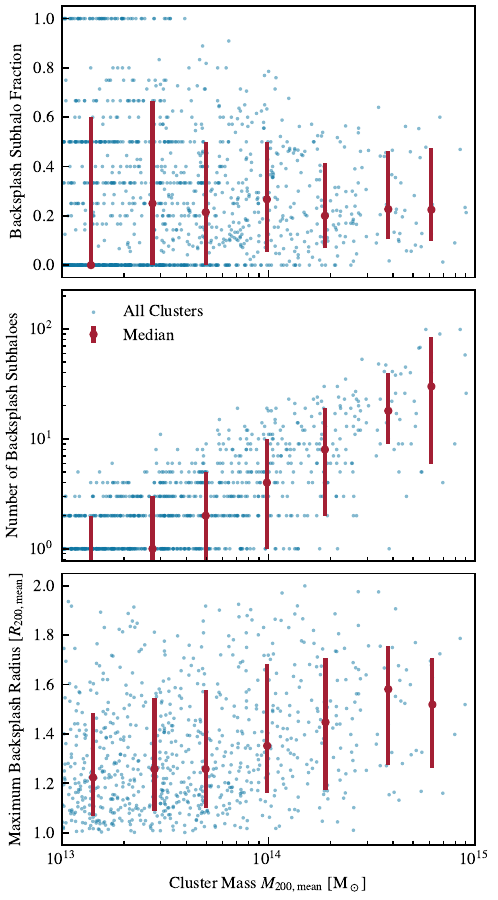}
    \caption{Shows properties of backsplash galaxies as a function of host
    cluster mass. In all panels, the blue dots represent the position of a
    single cluster in the plane, with the red error bars showing the 16th, 50th,
    and 84th percentile ranges (bottom, point, top) for clusters binned in 8
    equally log-spaced bins between $10^{13} \leq M_{\rm 200, mean} / {\rm
    M}_\odot \leq 10^{15}$. \emph{Top panel}: The fraction of subhaloes in the
    range $1.0 < R / R_{\rm 200, mean} < 2.0$ that are classified as backsplash
    galaxies.  \emph{Middle panel}: The total number of backsplash galaxies
    within this range as a function of cluster mass. This only edges
    above a median of one in clusters with $M_{\rm 200, mean} > 10^{13.5} ~ {\rm
    M}_\odot$. \emph{Bottom panel}: The maximum radius at which a backsplash
    galaxy is found for all clusters.}
    \label{fig:backsplash_numbers}
\end{figure}

Fig. \ref{fig:backsplash_numbers} shows some basic properties of backsplash
galaxies as they scale with the cluster mass $M_{\rm 200, mean}$. The top panel
shows the fraction of all subhaloes within the range $1 < R/R_{\rm 200, mean} <
2$ that are backsplashers, which is consistenly around 30\% across the cluster
mass range. Our backsplash fraction is lower than previously studied
simulations, with \citet{Gill2005} finding a backsplash fraction of 50\%, and
\citet{Haggar2020} finding that the backsplash fraction is of order 60\%.
\citet{Haggar2020} uses significantly higher mass clusters ($5\times10^{14}<
M_{\rm 200, crit} / {\rm M}_\odot < 3 \times 10^{15}$) than this work, and apply
a much higher mass cut on their galaxies (with $M_{\rm *} > 10^{9.5}$ M$_\odot$).
Additionally, they find that the backsplash fraction depends on the relaxation
of the cluster, with the most relaxed clusters having backsplash fractions less
than 20\%. Our isolation and tracking criteria for our clusters, required as
they live in live, full volume simulations (as opposed to zoom-in simulations of
clusters like {\textsc{TheThreeHundred}} sample), likely selects more relaxed
clusters. As we are mainly interested in the properties of the backsplash
galaxies themselves, rather than their host clusters, we defer further
investigation to potential future work.

The central panel shows that below cluster masses of $M_{\rm 200,
mean} = 10^{14}$ M$_\odot$, a very low fraction of clusters have any backsplash
galaxies at all, and the number of backsplash subhaloes scales roughly linearly
with cluster mass. This is primarily due to two factors: their smaller values of
$R_{\rm 200, mean}$ leading to a smaller volume within which backsplash galaxies
may reside, and their lower density environments.

Finally, the bottom panel of Fig. \ref{fig:backsplash_numbers} shows the
distribution of the maximal radii at which backsplash subhaloes are found
($R_{\rm sub, max}$ in Fig. \ref{fig:fancyimage}). In this panel, clusters are
not included if they do not have any backsplash subhaloes. We find that the
maximal radius, relative to $R_{\rm 200, mean}$, increases with cluster mass.
This is particularly notable as the splashback radius $R_{\rm sp}$ has been
shown by many authors, in particular \citet{O'Neil2021} for this sample, to
decrease with cluster mass.  Typical values of the splashback radius are roughly
$R_{\rm sp} / R_{\rm 200, mean} \approx 1.1-1.2$, as measured based upon the
dark matter profiles, trending down with cluster mass.  That there are many
galaxies outside of the typical splashback radius for these clusters is not
particularly surprising; the splashback radius should be the point at which
backsplash galaxies and infalling galaxies are at rough equipartition (assuming
galaxy number density traces mass density). We attribute the increase in maximal
backsplash galaxy radius with mass to the better sampling that these haloes
have, with them simply being more likely to find a galaxy out at this radius.

\begin{figure}
    \centering
    \includegraphics{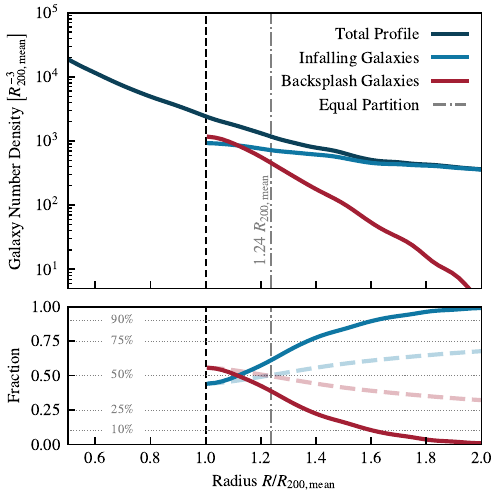}
    \caption{\emph{Top panel}: The number density profile of all galaxies with $M_* >
    10^8 ~ {\rm M}_\odot$ in our sample, stacked for all clusters in the sample.
    The dark blue line shows the profile of all galaxies, with the vertical
    black dashed line deliniating the point at which backsplash galaxies are
    classifed ($R_{\rm 200, mean}$). Outside this radius, the infalling galaxies
    (light blue) and backsplash galaxies (red) have their profiles shown
    separately, with the infalling galaxies tracing an almost constant density,
    and the density of the backsplash galaxies falling off rapidly outside of
    the cluster. \emph{Bottom Panel}: The volume-weighted fraction of backsplash
    and infalling galaxies as a function of radius. Horizontal dotted lines are
    shown at common percentile ranges to guide the eye. By $1.6R_{\rm 200,
    mean}$ fewer than 10\% of galaxies are backsplash, making attemtps to
    classify galaxies outside this radius infeasible. The dashed, lighter, lines
    shown in the background show the cumulative fraction of galaxies in each
    classification out to this radius (i.e. the fraction of galaxies with radius
    $r < R$ that are backsplash or infalling galaxies).  The equal partition
    between these two classifications is reached at $1.24 R_{\rm 200, mean}$
    (grey dot-dash line)}
    \label{fig:number_density_profile}
\end{figure}

Fig. \ref{fig:number_density_profile} shows the radial number density profile of
galaxies, stacked across all clusters in our sample, and resacled by $R_{\rm
200, mean}$. Even by $R = 1.1R_{\rm 200, mean}$, galaxies that are infalling
become more abundant, indicating just how rare backsplash galaxies are.  We
indicate that within the annulus of $1.0 < R / R_{\rm 200, mean} < 1.24$ the
density of backsplash galaxies and infalling galaxies are equal.
\citet{Bakels2021} finds that the radius at which backsplashing and infalling
subhaloes have equal density is at around $R / R_{\rm 200, crit} \approx 1.8$
(their Fig. 3)\footnote{We remind readers that ratios of $R_{\rm 200, mean} /
R_{\rm 200, crit}\approx1.65$ are typical for cluster haloes.}, similar to our
$R / R_{\rm 200, mean} \approx 1.1$,  making our results consistent despite
their more complex tracking procedure. We find little evolution in the equipartition
radius with host cluster mass, with it placed at $R / R_{\rm 200, mean} = 1.26$
and $1.24$ for clusters in mass ranges $10^{13} < M_{\rm 200, mean} / {\rm M}_\odot
< 10^{14}$ and $10^{14} < M_{\rm 200, mean} / {\rm M}_\odot < 10^{15}$ respectively,
and hence only show the stacked profile across our entire cluster sample.

Fig. \ref{fig:number_density_profile} shows that the infalling component has a
near-constant density across the radial range, with the backsplash (orbital)
component falling off rapidly with radius. This suggests that looking for 
galaxies bound to the cluster much further away than $R / R_{\rm 200,mean}
\approx 1.5$ is unlikely to yield a large number of results. Further, it
suggests that galaxies found this far away are unlikely to have been processed
by the cluster previously, meaning that in all liklehood their properties are
determined by their internal evolution rather than their environment.

\begin{figure}
    \centering
    \includegraphics{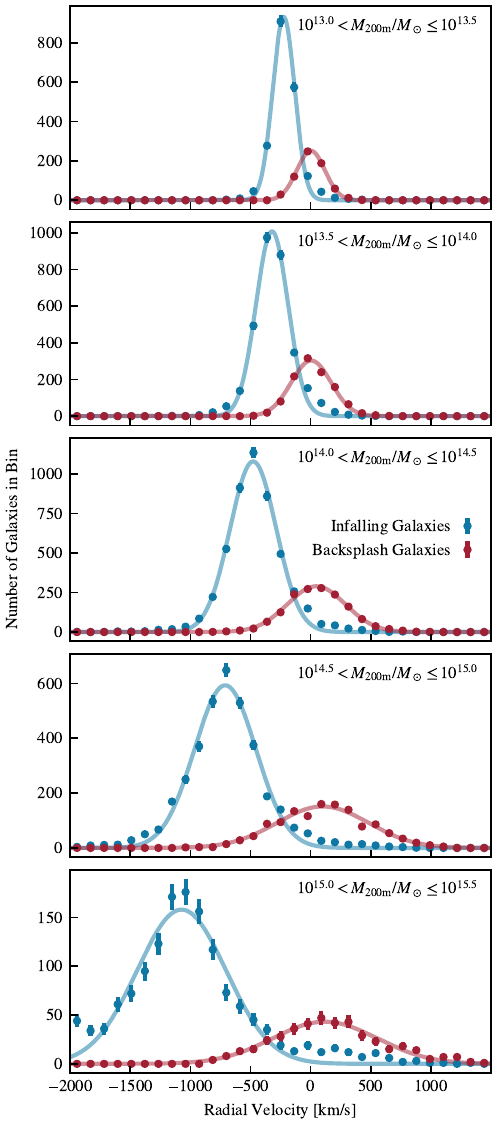}
    \caption{The radial velocity distributions for galaxies within $1 < R /
    R_{\rm 200, mean} < 2$, split by infalling/backsplash category (blue and red
    points, respectively), split by halo mass (different panels). Radial
    velocities are binned in 32 equally spaced linear bins between $-2000 ~ {\rm
    km} ~ {\rm s}^{-1} \leq v_{\rm r} \leq 1500 ~ {\rm km} ~ {\rm s}^{-1}$. The
    background lines show gaussian fits to the binned data to guide the eye.
    Infalling galaxies have, almost universally, a negative radial velocity (as
    expected). Backsplash galaxies have a gaussian distribution of radial
    velocities centred around, approximately, zero; at higher masses, though,
    the backsplash galaxies are biased to having a mean velocity greater than
    zero (i.e. there are more galaxies on their way back out of the cluster than
    falling back in). Notably the results remain qualitatively consistent (other
    than the relative abundances) when only considering galaxies within $R <
    1.24 R_{\rm 200,mean}$.}
    \label{fig:radialvel}
\end{figure}

In Fig. \ref{fig:radialvel} we show the radial velocity distributions of the
cluster galaxies in our sample, split by host cluster halo mass. Here we
calculate radial velocity disregarding the negligeable Hubble flow component (of
order 100 km s$^{-1}$) to better understand the gravitational dynamics at play
in clusters. This becomes non-negligeable for the highest mass clusters that
host galaxies at the largest radii in the lower panels, meaning that there is an
offset in the central velocity for backsplash galaxies. The radial velocity is
calculated in the rest frame of the cluster and
is given by
\begin{equation}
    v_{\rm r} = \frac{(\vec{v}_{\rm gal} - \vec{v}_{\rm clu}) \cdot (\vec{x}_{\rm gal} - \vec{x}_{\rm clu})}{|\vec{x}_{\rm gal} - \vec{x}_{\rm clu}|},
    \label{eqn:vr}
\end{equation}
where here $\vec{v}$ and $\vec{x}$ are the physical velocity and position of the
cluster and galaxy (subscripts clu and gal respectively) relative to the
simulation box.

We see that the velocities of backsplash galaxies generally follow a normal
distribution around zero, with roughly equal numbers of galaxies on their way
out ($v_{\rm r}> 0$) as in ($v_{\rm r} < 0$) to the cluster. The infalling
galaxies, on the other hand, almost universally are falling into the cluster
($v_{\rm r} < 0$, note that we did not make a velocity-based cut in our
criterion for infalling galaxies, just that they must be near a cluster and not
have been in the cluster previously). In almost all cases, the absolute value of
the velocity for backsplash galaxies is significantly lower than the velocity of
infalling galaxies.  The infalling galaxies have a mean velocity that scales
with the cluster mass, with higher mass clusters having significantly faster
infall speeds ( $v_{\rm r}\approx -250$ km s$^{-1}$ for $10^{13.0} < M_{\rm 200, mean}
/ {\rm M}_\odot \leq  10^{13.5}$ clusters, versus $v_{\rm r} \approx -1100$ km s$^{-1}$
for $10^{15.0} < M_{\rm 200, mean} / {\rm M}_\odot \leq  10^{15.5}$ clusters).
The higher infall velocities of galaxies in high mass clusters is a potential
explanation for the increase of $R_{\rm sub, max}$ with mass seen in the bottom
panel of Fig. \ref{fig:backsplash_numbers}.

It is hence clear why it is common to use phase-space information to separate
orbiting and infalling galaxies, as in \citet{Farid2022}. It is possible to
easily classify a galaxy into the infalling or backsplash bins based upon
the radial velocity alone. Unfortunately, measuring the radial velocity of
galaxies realtive to their cluster is extremely difficult in observations.
Velocities measured in observations cannot decompose components
into three dimensions. Only velocities along the line of sight can be measured,
with redshift offsets, but these provide little information about the radial 
velocity of the galaxy (this is only one component of three). Thus, given a
line of sight velocity of zero, it is unclear whether this is an infalling 
galaxy with a high $v_r$, just tangential to the viewer, or a backsplash galaxy
with $v_r = 0$ at apocentre.

As such, classifying a galaxy based upon its intrinsic properties (e.g.
luminosity) so that it can be targeted for confirmatory follow-up and further
study is desireable. 

\begin{figure*}
    \centering
    \includegraphics{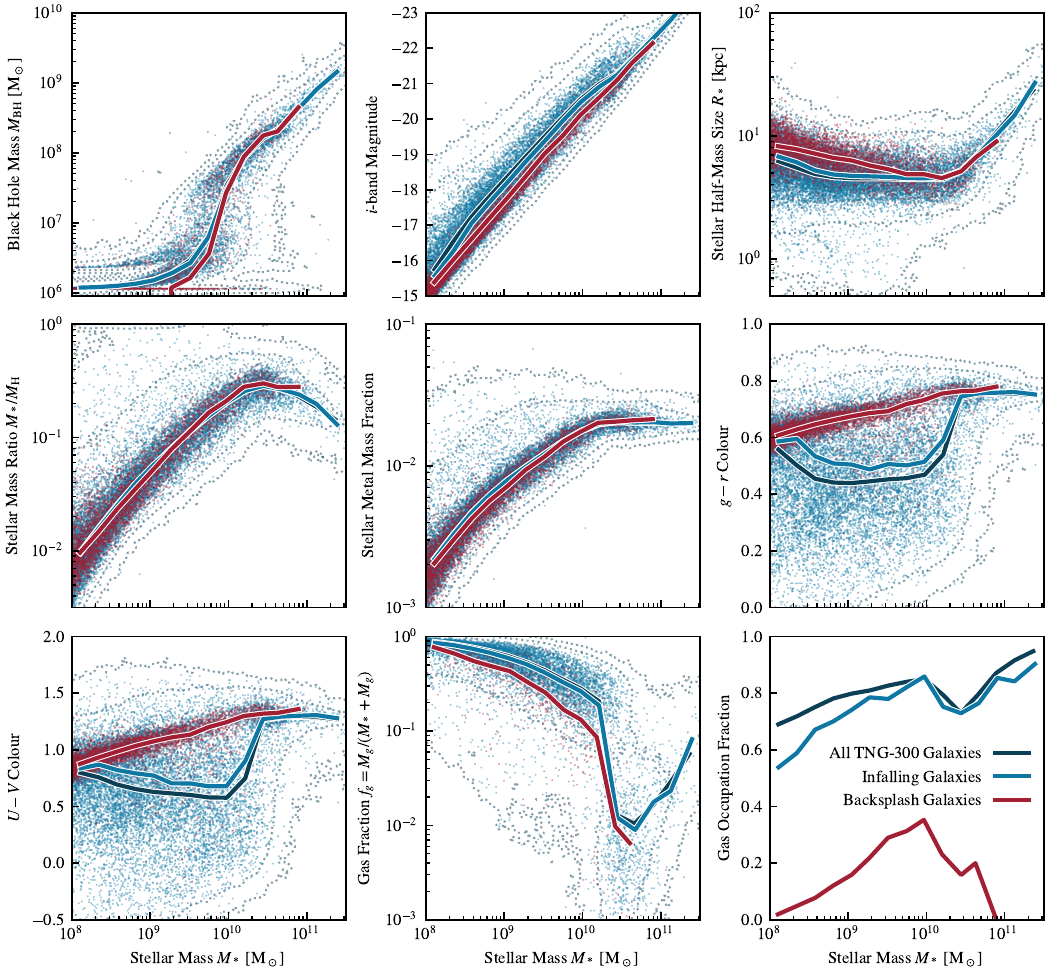}
    \caption{A collection of common galaxy scaling relations, split into three
    groups: backsplash galaxies (red points and line), infalling galaxies (light
    blue points and line), and all TNG300 galaxies (dotted contours and dark
    blue line). Lines show the median binned in 16 equally log-spaced bins
    across the stellar mass range shown, and the points show all galaxies. All
    scaling relations are shown as a function of galaxy stellar mass, here
    defined to be the mass contained within twice the stellar half-mass radius.
    Here we see that backsplash galaxies have lower mass black holes, high mass
    to light ratios, and are generally redder, larger, and have significantly
    lower gas occupation fractions than their infalling counterparts.}
    \label{fig:scalingrelations}
\end{figure*}

In Fig. \ref{fig:scalingrelations} we show a number of galaxy scaling relations
for the galaxies in our sample, split into three categories: backsplash (red),
infalling galaxies (blue), and for comparison purposes, all galaxies in the
TNG300 simulation (dark blue and background contours). Each line shows the
median relation for these galaxies, binned in 16 equally log-spaced bins across
the mass range shown in the figure. In the background we show scatter plots,
coloured appropriately, for both the backsplash and infalling galaxies, to give
some sense of scatter in the relations.

The general picture from these scaling relations is that backsplash galaxies are
redder and less gas-rich than their infalling counterparts. All scaling
relations are shown against the final, $z=0$, stellar mass of the galaxies (i.e.
what would be observed), rather than the pre-infall mass of the galaxies. We
choose to do this as we aim to classify galaxies based upon their properties,
and pre-infall properties of galaxies cannot be observed.  Generally,
infalling galaxies are consistent with the entire TNG300 population.

In the top left panel we show the black hole mass, which is the sum of all black
hole masses that are bound to the substructure. We see that, below $M_* \approx 2
\times 10^9$ M$_\odot$, backsplash galaxies (as the median) have no black
holes, but above $M_* \approx 10^{10}$ M$_\odot$ backsplash galaxies follow the same
trend as infalling galaxies. We dedicate \S \ref{sec:blackholes} to the investigation
of this surprising property of backsplash galaxies in TNG300.

The top centre panel shows the $i$-band (typically the brightest out of the
eight bands in the TNG catalogue) magnitude of galaxies as a function of their
stellar mass. Backsplash galaxies live in a very thin band around the median,
with far (0.5dex) higher (dimmer) magnitudes then their infalling counterparts,
up to $M_* \approx 10^{10.5}$ M$_\odot$. This higher mass-to-light ratio for
backsplash galaxies suggests that their stars are redder and older, and has been
observed in prior studies \citep{Knebe2011}. At $M_*
>10^{10.5}$ M$_\odot$, AGN play an important role in shutting down star formation
in the galaxies, quenching them internally (rather than through external processes
such as ram-pressure stripping), reducing their mass-to-light ratio \citep{Donnari2019}.

The top right panel shows the half-mass stellar size of galaxies. We see that
backsplash galaxies are baised towards having higher sizes than their infalling
counterparts at $M_* < 10^{10}$ M$_\odot$, though there is a high degree of
overlap in the two size distributions. We dedicate \S \ref{sec:sizes} to the
investigation of size differences between backsplash and infalling galaxies.

In the centre left panel we show the stellar mass ratio $M_* / M_{\rm H}$, the
fraction of bound mass to the galaxy that is in the stellar component.
Backsplash galaxies and infalling galaxies show a similar stellar mass ratio
across the mass range to their infalling counterparts, suggesting that
backsplash galaxies have not preferentially lost or gained stellar mass, and
that differences in their stellar properties are due to rearangement of stars
and star formation either temporally (e.g. quenching) or spatially (e.g. tidal
heating).

The centre panel shows the stellar metal mass fraction (frequently referred to
in simulations as `metallicity'), which is the fraction of the stellar mass in
the galaxy that is made up of elements other than hydrogen and helium. Below
$M_* < 10^{10}$ M$_\odot$ backsplash galaxies have a slightly (0.05--0.1 dex)
lower metallicity than infalling galaxies, and have far fewer high metallicity
galaxies.

The $g-r$ and $U-V$ colours of the galaxies are shown in the centre right
and lower left panels, confirming the suspicions from Fig. \ref{fig:fancyimage} that
backsplash galaxies are on average redder than infalling galaxies. The backsplash
galaxies live in the well-studied red sequence of quenched galaxies, indicating
that they have been quenched for a long time ($> 1$ Gyr). Though there are many
quenched infalling galaxies, there is a significant active population with colour
$U-V\approx 0.5$. At the high mass end, $M_* > 2\times10^{10}$ M$_\odot$, we again
see that the infalling galaxies become quenched (coinciding with high black
hole masses, and high mass-to-light ratios in the upper panels), further indicating
that AGN feedback rapidly quenches galaxies at this mass.

In the final two (lower centre and right) panels, we consider the gas fraction of
the galaxies. The gas fraction,
\begin{equation}
    f_{\rm g} = \frac{M_{\rm g}}{M_* + M_{\rm g}},
\end{equation}
is shown for galaxies with at least one bound gas particle in the lower centre
panel. Even in galaxies that retain gas, the gas fractions of backsplash
galaxies are significantly lower than their infalling counterparts. Another
indicator of the significance of AGN feedback can be seen at $M_* \approx
2\times10^{10}$ M$_\odot$, where all galaxies see a sharp downturn in gas
fraction thanks to the evacuating effect of these objects.

In the bottom right panel we show the gas occupation fraction of galaxies as a
function of their stellar mass. The gas occupation fraction is defined as the
fraction of galaxies in a bin that have a gas mass greater than zero, i.e. they
contain bound gas particles. Notably, backsplash galaxies have significantly
lower (3--4 times) gas occupation fractions than infalling galaxies. This staggering
difference is even higher when considering clusters with $M_{\rm 200, mean} >
10^{14}$ M$_\odot$, (not shown for brevity) where nearly all backsplashers have
no gas. Such low gas occupation fractions likely originate from strong ram
pressure stripping of galaxies as they pass through the cluster
\citep{Simpson2018}.

\section{Identifying Critical Properties}
\label{sec:classifier}

In this section, we use a decision tree classifier to identify backsplash
galaxies based upon their intrinsic properties. We choose a decision tree
classifier as it is highly interpretable; the outputs from the model are simply
bisections in each parameter space, ranked by their importance. Through the
decision tree, we can identify (in a relatively unbaised way) which features of
galaxies can be used to classify them as backsplash or infalling galaxies,
before moving on to the physical interpretation of these features. We note that
the main reason for employing the machine learning model in this context is not
necessarily to enable the use of an off-the-shelf model for using intrinsic
properties to identify backsplash galaxies, as the specific normalisations of
such quantities (stellar mass, black hole mass, galaxy colour, etc.) are usually
not entirely converged within the simulation and modelling errors may lead to
offsets to observational data \citep{Schaye2015, Pillepich2018}. Instead, we
wish to use the model to tell us which, out of all available properties, are
those critical to bisecting the population of backsplashing and infalling
galaxies.

\subsection{Training the decision tree}

\begin{table*}
    \centering
    \begin{tabular}{l|l|c}
    {\bf Galaxy Property} & {\bf Aperture} & {\bf Log?} \\
    \hline
        Total black hole mass & Total & $\checkmark$ \\ 
        Total black hole mass accretion rate & Total & $\checkmark$  \\ 
        Magnetic field strength in disk & Total & $\checkmark$  \\ 
        Magnetic field strength in halo & Total & $\checkmark$  \\ 
        Center of mass & Total & $\times$  \\ 
        Gas metal mass fractions for each species & Twice half-mass radius, half-mass radius, max radius, SF gas only & $\checkmark$  \\ 
        Gas total metallicity & Twice half-mass radius, half-mass radius, max radius, SF gas only & $\checkmark$  \\ 
        Galaxy half-mass radius (per species) & Total & $\checkmark$ \\ 
        Galaxy half-mass radius (total) & Total & $\checkmark$ \\ 
        Number of particles (per species) & Total & $\checkmark$ \\ 
        Number of particles (total) & Total & $\checkmark$ \\ 
        Total mass & Total, half-mass radius (per species, total), twice half-mass radius, max radius & $\checkmark$ \\ 
        Subhalo position in box & Total & $\checkmark$ \\ 
        Subhalo instantaneous SFR & Total, twice half-mass radius, max radius & $\checkmark$ \\ 
        Subhalo spin parameter $\kappa$ & Total & $\checkmark$ \\ 
        Star metal mass fractions for each species & Twice half-mass radius, half-mass radius, max radius & $\checkmark$ \\ 
        Star total metallicity & Twice half-mass radius, half-mass radius, max radius, SF gas only & $\checkmark$ \\ 
        Magnitude in $U$, $B$, $V$, $K$, $g$, $r$, $i$, $z$ bands & Total & $\times$ \\ 
        Total mass within K-band half-light radius & Total & $\checkmark$ \\ 
        Half-light radius in K band & Total & $\checkmark$ \\ 
        1D velocity dispersion of all member particles & Total & $\checkmark$ \\ 
        Maximum value of the spherically-averaged rotation curve $V_{\rm max}$ & Total & $\checkmark$ \\ 
        Radius at which $V_{\rm max}$ is achieved. & Total & $\checkmark$ \\ 
        Total mass currently decoupled as winds & Total & $\checkmark$ \\ 
        $U-V$ and $g-r$ colour & Total & $\times$ \\ 
    \hline
    {\bf Cluster Property} & {\bf Aperture} &  {\bf Log?} \\
    \hline
        Group centre of mass & Total & $\times$ \\
        Group gas metal mass fractions for each species & Total & $\checkmark$ \\
        Group gas metallicity & Total & $\checkmark$ \\
        Group star metal mass fractions for each species & Total & $\checkmark$ \\
        Group star metallicity & Total & $\checkmark$ \\ 
        Group aperture masses & $R_{\rm 500, crit}$, $R_{\rm 500, mean}$, $R_{\rm 200, crit}$, $R_{\rm 200, mean}$, $R_{\rm 200, tophat}$ & $\checkmark$ \\
        Group aperture radii & $R_{\rm 500, crit}$, $R_{\rm 500, mean}$, $R_{\rm 200, crit}$, $R_{\rm 200, mean}$, $R_{\rm 200, tophat}$ & $\checkmark$ \\
    \end{tabular}
    \caption{\emph{Top}: The galaxy properties used to train the decision tree. In the second column, the
    aperture within which this quantity is calculated is noted. Here, total refers to the total
    bound mass to this subhalo. In this column, half-mass radius refers to the \emph{stellar}
    half-mass radius. \emph{Bottom}: The cluster properties associated with each galaxy. These
    are the properties of the cluster that the galaxy has backsplashed from or is falling into.
    The final column for both shows whether or not the value is pre-scaled logarithmically
    before its use in the decision tree. Note that this does not affect results, as the decision tree
    simply performs a set of bisections, but it does aid in the inspection of the results.}
    \label{tab:properties}
\end{table*}

Once we have identified all backsplash galaxies and infalling galaxies within
the radius range $1 < R / R_{\rm 200, mean} < 2$, we assign all galaxies integer
labels (zero for infalling galaxies, and one for backsplash galaxies). Alongside
this, we collate all (reasonable) galaxy properties from the publicly available
TNG SubFind catalogues. These are shown in Table \ref{tab:properties}, with the
properties of their hosts that are also included in their description. Finally,
we include the current distance between the galaxy and cluster centre, $R_{\rm
gal}$ in units of the cluster $R_{\rm 200, mean}$.  Most properties are
logarithmically scaled ($R_{\rm gal}$ is not), as this eases in the
interpreation of the results from a physical standpoint. As the decision tree
framework does not accept any values that are infinity or not a number, we bound
all scaled values as a lower limit of -10. Typically when a property has a
physical value of zero (and hence logarithm of {\tt -inf}), the decision tree
uses a cut to check for occupation of this value (e.g. gas mass), so the
specific value chosen here does not impact our results.

The results in this section are based upon the clusters with mass $10^{14} <
M_{\rm 200, mean} / {\rm M}_{\odot} < 10^{15}$, to reduce complexities
associated with cluster evolution but to still leave a large sample of clusters
(215) and backsplash galaxies.  Additionally, we make a cut in galaxy stellar
mass $M_* > 10^8$ M$_\odot$ (10 particles), mainly to ensure that there are
enough particles such that the occupation of the subhalo is accurate. At low
masses, where there are only one or two particles in a subhalo, there may be
many similar subhaloes that are unoccupied simply by chance, as star formation
in the TNG model is a stochastic process \citep{Genel2019, Keller2019}.

To train the decision tree, we transform all associated features ($N_{\rm feat}$) with the
$N_{\rm gal}$ galaxies into a $N_{\rm feat} \times N_{\rm gal}$ matrix, with an associated
one dimensional array containing the integer labels. These two arrays are then shuffled, to
simplify training and validation splits.

We leverage the decision tree classifier object already implemented in the
{\textsc{python}} package {\textsc{Scikit-Learn}} \citep{scikit-learn}. When
training, we use the Gini impurity for the split criterion, the ``best'' split
strategy, a maximum tree depth of 6, and a minimal impurity decrease of 0.001.
Choosing instead to use the entropy information gain criterion, or larger tree
depths, do not qualitatively change our results. We choose the depth and minimum
impurity decrease to make the tree easier to visually inspect, and for performance
reasons.

To investigate the accuracy of our model, we use a 80--20 validation and training
split. Our decision tree gets a 20\% cross-validation score of 89\%, meaning that when 
trained on a 80\% of the data, and tested on 20\%, we are able to correctly predict
the classification of a galaxy 89\% of the time. However, as we are mainly interested
in the ability of the model to predict whether or not something is a backsplasher, 
rather than the ability of the model to predict if a galaxy is infalling, we now
focus on our positivity and false positive rate. When considering only galaxies
that are backsplashers, we have a 20\% positivity rate of 78\%, meaning that we
mis-classify 22\% of backsplash galaxies as infalling galaxies. However, we only
have a 7\% false-positive rate, meaning only 7\% of the galaxies predicted by
the model to be backsplash galaxies are actually infalling. The accuracy of our
decision tree model is comparable to other recent studies that additionally
employ phase-space information, with \citet{Farid2022} able to achieve an 83\%
classification accuracy when using phase-space information and sSFR with a
random forest model. The hard classification model\footnote{Note that `hard' in
this context implies that the model must make a choice to classify an object
within a gorup, as opposed to `soft' classification that assigns probabilities
of belonging to a given group to each object.} in \citet{Farid2022} is 
designed to separate infalling galaxies and those within the cluster, however,
and does not explicitly classify backsplashers. Their hard classification model
cannot accurately predict the existence of bound structures to the cluster
in the range $R > R_{\rm 200, mean}$.

Such a high level of accuracy is surprising given the large level of overlap
between the backsplash and infalling population (see Fig.
\ref{fig:scalingrelations}). For instance, making a cut based on simple
population statistics, with $R / R_{\rm 200, mean} < 1.2$, and a red sequence
cut in $g-r$ magnitude, returns a positivity rate of under 30\% (though with
only a 3\% false-positive rate owing to the markedly lower number of galaxies
classified as backsplashers).

\subsection{Interpreting the model}

\begin{figure}
    \centering
    \includegraphics{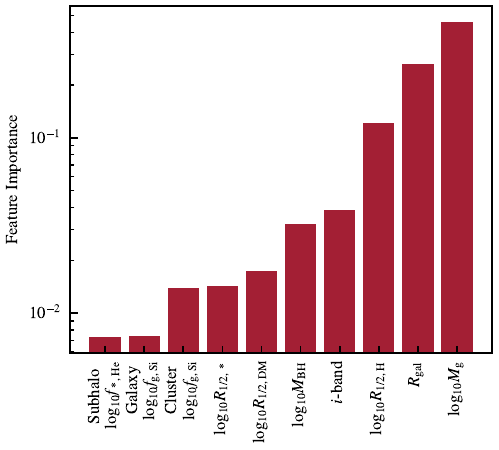}
    \caption{Gini importances of the top ten highest ranked features in the TNG300
    datasets, when classifying backsplash galaxies. A higher importance score
    corresponds to a higher level of information gain when bisecting this property.
    Descriptions of the symbols used here are available in the text.}
    \label{fig:feature_importance}
\end{figure}

\begin{figure*}
    \centering
    \includegraphics{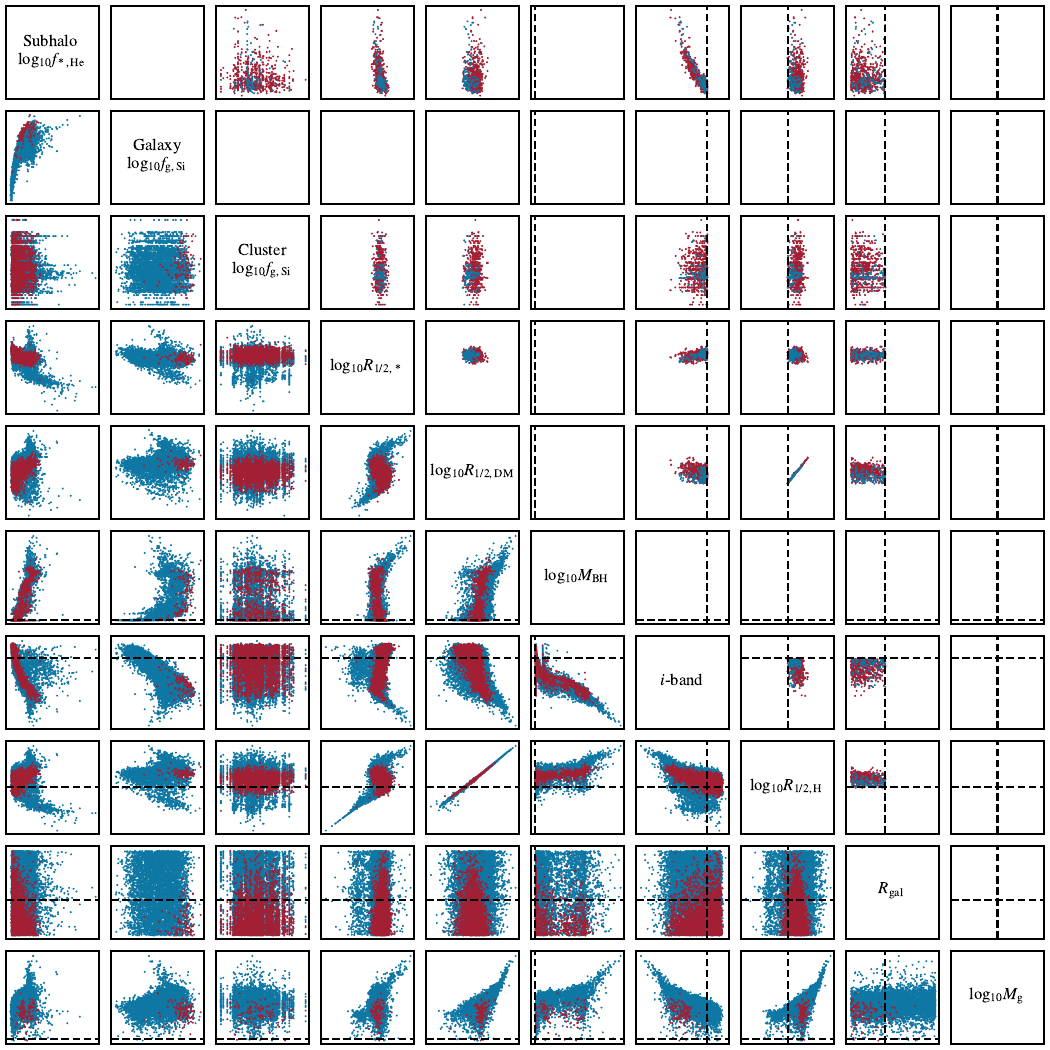}
    \caption{The highest ranked features from Fig. \ref{fig:feature_importance} scattered
    against each other. The bottom left triangle shows all galaxies in the sample, with
    blue points representing infalling galaxies and red points representing backsplash
    galaxies. The dashed lines show the cuts identified by the decision tree based on
    the top five most important features, with the top right panels showing the resulting
    parameter space after the cuts have been applied to find the largest contiguous volume
    of backsplash galaxies. This time, the blue points are plotted
    on top of the red points to highlight false positive detections. In this region, we find
    603 backsplash galaxies with a false-positive rate of 15\%.}
    \label{fig:scatter}
\end{figure*}

That we can classify galaxies as backsplashers from their intrinsic properties
without detailed kinematic information is interesting, but not as interesting
as answering \emph{why} we can classify them. Fig. \ref{fig:feature_importance} shows
the feature (Gini) importances for the top ten features in the model for bisecting
the infalling and backsplash galaxy populations. Their correlations are also
shown in Fig. \ref{fig:scatter}, where backsplash galaxies are shown in red, and
infalling galaxies in blue. In this figure, dashed black lines show the bisections that
the decision tree predicts are the most likely to segment the two populations, with the
series of cuts in the tree shown for the leaf node with the highest number of backsplash
galaxies. The top right panels show the same projections but with the cuts applied to
the whole population.

The most important feature in the entire model is $\log_{10} M_g$, the gas mass
of the galaxy. We can see that this feature is used for occupation testing by
the model, with the majority of backsplash galaxies (see the bottom right panel
of Fig. \ref{fig:scalingrelations}) containing zero gas. This is even more stark
when applied to the more massive clusters, with gas occupation fractions of
backsplash galaxies reducing as the clusters get more massive due to increased
ram pressure stripping.

For the second most important feature, it is fair to say that the decision tree
has `learned' about the geometry of the problem (Fig. \ref{fig:number_density_profile}),
with the liklehood of a galaxy being a backsplasher dropping rapidly with distance
from the cluster centre. The decision tree places a cut when just over 50\% of
the galaxies are likely to be infalling ($R_{\rm gal} / R_{\rm 200, mean} = 1.41$).
This is clear to see as well in Fig. \ref{fig:scatter}, where the frequency
of backsplash galaxies clearly drops off with increasing $R_{\rm gal}$.

Next, there are three measures of galaxy size. In decreasing order of
importance, the model finds that $\log_{10} R_{\rm 1/2, H}$ (total bound
half-mass radius), $\log_{10} R_{\rm 1/2, DM}$ (dark matter half-mass radius),
and $\log_{10} R_{\rm 1/2, *}$ (stellar half-mass radius) are useful indicators
of backsplash status. As this topic is somewhat complex, we dedicate an entire
section of the paper to it (\S \ref{sec:sizes}), but briefly we see that
backsplash galaxies tend to have larger stellar half-mass sizes, and larger (or
smaller) dark matter half-mass sizes for higher mass (lower mass), than their
infalling counterparts.

The model finds that a cut in $i$-band magnitude is useful in segmenting the two
populations, but it chooses a very low luminosity cut (with $i < -16.7$). In
Fig. \ref{fig:scalingrelations}, we see that at these low magnitudes the overlap
between the infalling and backsplash populations becomes higher, making
segmenting them based upon this property troublesome. Although the prescence of
this cut is useful, it is likely to be more numerically motivated, as at this
luminosity (and hence mass) photometry is not particularly well resolved, with
the star formation history of the galaxy being traced by around 10 stellar
particles. With increasing resolution, the turn down of the $i/\log M_*$
relation seen in Fig. \ref{fig:scalingrelations} at the lowest luminosities
would likely disappear, but it is notable that the model picks up numerical
differences between populations, as well as physical differences,
indiscriminately.

In a similar fashion to the gas mass of galaxies, the model uses the black
hole mass as an occupation check, with it placing a cut in the black hole
mass $\log_{10} M_{\rm BH}$ below the seed mass of black holes in the simulation.
This is why in the top right of Fig. \ref{fig:scatter}, the black
hole mass panels contain zero points; the model has selected only galaxies
that currently have zero black hole mass. As with the sizes, we dedicate
a section (\S \ref{sec:blackholes}) to the investigation of this.

% {\color{red} I don't really have a good understanding of why this is
% at this point.}
Finally, the model finds that some species fractions contribute to the ability
to bisect the two populations. First, the cluster silicon gas fraction
($\log_{10} f_{\rm g, Si}$), which is the mean silicon gas fraction within all
gas particles bound to the cluster associated with the galaxies. Secondly, we
find that, in galaxies that retain their gas, their gas silicon abundance is a
useful criterion to bisect on. Finally, the fraction of galaxy stellar mass that
is helium ($\log_{10} f_{\rm *, He}$) contributes to the ability to bisect the
two populations. \citet{Vogelsberger2018} and \citet{Barnes2018} investigated
clusters in the TNG300 simulation, and showed that cool-core clusters, with
relaxed morphologies, have significantly higher levels of enrichment in their
cores. This may suggest that higher cluster enrichment levels may correlate with
different tidal interactions between galaxies and the cluster, but we see no
significant correlation between $\log_{10} R_{\rm 1/2, DM}$ and $\log_{10}
f_{\rm g, Si}$ in Fig. \ref{fig:scatter}. It is additionally clear that further
cuts on the cluster silicon abundance and galaxy helium abundance would be able
to separate out a cleaner sample of backsplash galaxies, but along with the low
importance indicated by the decision tree it is likely that this is simply by
chance.
\section{Understanding Galaxy Sizes}
\label{sec:sizes}

In Fig. \ref{fig:feature_importance}, we saw that some of the leading-order
features separating backsplash and infalling galaxies were there sizes in
various mass types (total mass, dark matter mass, and stellar mass).  These
splits were used downstream of the occupation cut in gas mass, with star
formation rates in the halo catalogues computed instantaneously from the bound
gas (and hence all of the backsplash galaxies were passive).  This means that
these features are useful irrespective of active/passive status of galaxies as
seen by the model.

\begin{figure}
    \centering
    \includegraphics{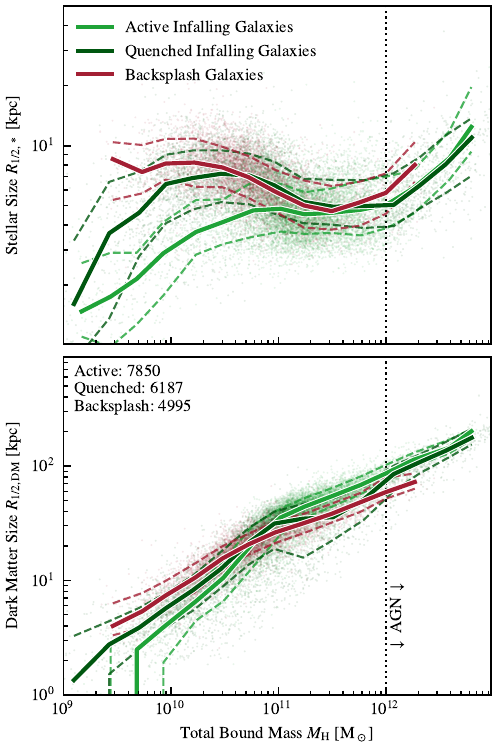}
    \caption{Mass-size relations for backsplash galaxies (red) and infalling
    galaxies split into quenched (dark green) and active (light green) populations
    by specific star formation rate ${\rm sSFR} = 10^{-11}$ year$^{-1}$. In the
    background the scatter shows all galaxies in the sample, with the solid
    lines representing the median in 16 equally log-spaced bins, and dashed
    lines showing the 16--84 percentile range. The top panel shows the
    stellar half-mass radius, and the bottom panel shows the dark matter
    half-mass radius. The dotted vertical line shows the total bound mass at which
    AGN feedback becomes significant, leading to a noticeable up-tick in the
    stellar sizes. The numbers in the top left of the bottom panel show how
    many galaxies are in each cut below the point at which AGN feedback becomes
    significant.}
    \label{fig:masssize}
\end{figure}

In Fig. \ref{fig:masssize} we show the relationship between total bound mass of
galaxies and their stellar and dark matter half-mass sizes (top and bottom
panels respectively). These relationships are split by their active and passive
nature for infalling galaxies, with backsplash galaxies grouped together (the
quenched fraction of backsplash galaxies is 94\% so splitting these makes little
sense). Here, we take the active/passive threshold to be a specific star
formation rate ${\rm sSFR} = {\rm SFR} / M_* =10^{-11}$ year$^{-1}$, computed from the
instantaneous gas star formation rate. Only galaxies with at least $M_* > 10^8$
M$_\odot$ are included in this analysis, but all clusters are included.

As in \citet{Genel2018}, we find that quenched galaxies have significantly
larger stellar sizes than their active counterparts. At the lowest masses
($M_{\rm H} < 10^{11}$ M$_\odot$), backsplash galaxies show even larger stellar
sizes than their quenched counterparts. 

In \citet{Genel2018} one of the main explanations for low-mass galaxies being
large is that they have experienced significant tidal heating, as the sample of
low mass galaxies are almost entirely satellites. Here, we are able to show
that `field' galaxies (in this context, these are galaxies that have not been
processed by the cluster yet, so infalling galaxies) that are quenched have
experienced similar size growth at low masses as the general quenched population,
suggesting a different explanation is warranted.

At the high-mass end of the stellar sizes, we see that backsplash galaxies grow
their sizes faster (at lower masses) than infalling galaxies, but there are few
backsplash galaxies of this mass and as such this result is tenuous. At the
highest masses, $M_{\rm H} > 10^{12}$ M$_\odot$, quenched and active galaxies
have similar sizes when separated by instantaneous specific star formation rate.
This similarity is likely due to the method of quenching; these galaxies are
mainly quenched by their central AGN periodically, going through phases of
feedback and gas evacuation, followed by fresh accretion and cooling leading
to increases in star formation. Here the quenched and active separations only show
different epochs of the evolution of these galaxies.

The dark matter sizes of galaxies are generally more similar, with a notable
kink in the infalling galaxy sizes at $M_{\rm H} = 10^{11}$ M$_\odot$ that is
not present in the backsplash galaxies. This kink is likely caused by the
introduction of black holes to the haloes, which occurs at total FoF halo masses
of $7.4 \times 10^{10}$ M$_\odot$ in the TNG model, with AGN feedback having
been shown to decrease the central density of dark matter in simulations
\citep{Duffy2010}.

At masses below the kink, backsplash galaxies are the largest, but this trend is
reversed after the kink where they are the smallest.  This indicates that at low
masses, similar processes are governing the size differences (i.e. low mass
galaxies have large stellar and dark matter sizes for the same reason) but at
higher masses the stellar component and dark matter components are affected by
different processes. In addition, at masses below $M_{\rm H} < 10^{11}$
M$_\odot$, stellar and dark matter half-mass sizes are comparable, of order 10
kpc. At higher masses, the dark matter half-mass size is significantly larger,
breaking the scales between the dark matter and stellar components.

\begin{figure}
    \centering
    \includegraphics{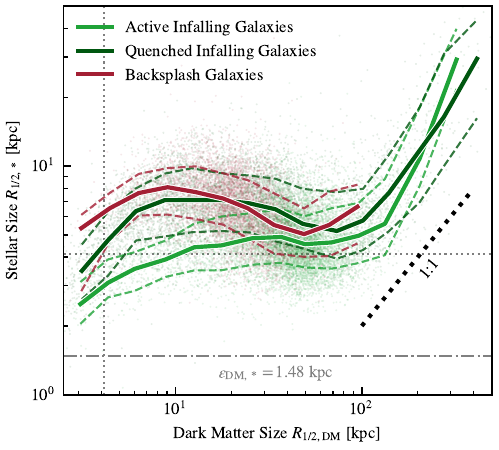}
    \caption{The two galaxy size measures from Fig. \ref{fig:masssize} shown
    against each other, with the same line styles and colours. In the background
    we show the 1:1 relation as a thick dotted black line, and for comparison
    purposes show the physical softening length at $z=0$ in the TNG300
    simulation of 1.48 kpc as a dot-dash grey line. We show, similarly to
    \citet{Genel2018}, that quenched galaxies show larger stellar sizes on
    average even for a fixed halo size.  The thin background dotted grey lines
    show the scale of $2.8\epsilon$, the characteristic scale identified in
    \citet{Campbell2017} at which sizes begin to flatten.}
    \label{fig:size_size}
\end{figure}

In Fig. \ref{fig:size_size} we show the two measures of galaxy size against each
other, alongside the softening scale. \citet{Campbell2017} found that galaxy
stellar sizes flatten at a charachteristic scale (2.8 times the softening length
$\epsilon_{\rm DM, *}$, which for TNG300 is 1.48 kpc at redshift $z=0$), and
this appears to be true within TNG300, also. However, whilst the active galaxy
sizes saturate at around 4.2 kpc, as predicted by \citet{Campbell2017}, our
quenched galaxy sizes (and backsplash sizes) rise above this level for the
lowest masses. As TNG300 haloes and galaxies grow, they begin to revert to
the 1:1 relation (at dark matter sizes above 100 kpc).

\citet{Ludlow2019a}, \citet{Ludlow2019b}, and \citet{Ludlow2020} discuss origins
of this potentially spurious size growth in galaxies, and suggest that
its origins lie in 2-body scattering between different particle species (in this case,
dark matter and stars) with different masses. In this scenario, energy is transferred
from the more massive component (dark matter) to the less massive component (stars),
causing the dark matter to be centrally concentrated within the halo and the
stellar component to be dynamically spread. Galaxies that are artificially inflated
by 2-body scattering
are then expected to have more massive, cuspier, inner dark matter profiles,
and hence smaller dark matter sizes at a given mass.

Our relationship between $R_{\rm 1/2, DM}$ and $R_{\rm 1/2, *}$ is generally
flat at dark matter sizes below 100 kpc (i.e. it does not continue down the 1:1
relation), indicating that sizes are propped up in this case by softening
inflation, with a minimal stellar (and dark matter) size set by $2.8 \epsilon
\approx 4.5$ kpc. There is clearly more to this, however, with there being
significant differences between active and passive infalling galaxies, as well
as the backsplash population.

In TNG-50, where significantly higher resolution simulations are available (a
rough factor of 100 smaller particle masses), \citet{Pillepich2019} showed that
at the resolution of TNG300, (roughly TNG-50-3) the stellar half-mass sizes of
galaxies were not converged below $M_* < 10^{9.5}$ M$_\odot$. Higher resolution
generally gives smaller stellar sizes, consistent with the indications of
\citet{Ludlow2021} that this may be due to lower levels of dynamical heating due
to the lower dark matter masses used, or that the sizes are inflated due to
oversoftening (as the chosen softening frequently scales with resolution). It is
hence likely that 2-body scattering, as well as size inflation caused by
softening, cause a systematic trend in our galaxy sizes at low masses ($M_* <
10^{9.5}$ M$_\odot$) to become larger, meaning that the absolute sizes of our
low-mass galaxies are overestimated. This does not mean, however, that any
differences between the galaxy populations on these scales are entirely
numerically driven. In particular, that we see differences in dark matter sizes
at $M_{\rm H} < 10^{11}$ M$_\odot$ is reliable, as these sizes are both much
larger than the softening scale (see Fig. \ref{fig:size_size}) and generally
would be made smaller, not larger, by 2-body scattering.

\begin{figure}
    \centering
    \includegraphics{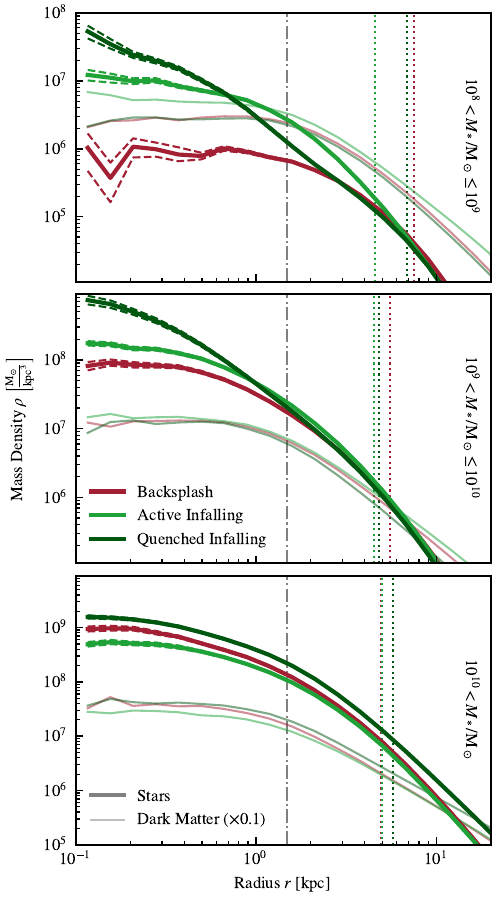}
    \caption{Mean stellar density profiles, split by final stellar mass
    (panels). Solid lines show the median, with the dashed lines showing the
    16--84 percentile range, of 1024 mean-stacked bootstraps (with replacement)
    of all individual profiles. The vertical grey dot-dash line shows the
    gravitational softening used for stellar and dark matter particles, and
    the dotted lines show the median stellar half-mass radius for galaxies
    in each mass bin of the corresponding colour. In the background the
    lighter, thinner, line show the dark matter density profile, multiplied
    by 0.1 for clarity.}
    \label{fig:stellarmassprofile}
\end{figure}

In Fig. \ref{fig:stellarmassprofile} we show the mean-stacked stellar mass and
dark matter mass profiles for the inner 20 kpc of all galaxies within each mass
range. These profiles include all bound matter, and are stacked for three
different mass ranges: $10^8 < M_* / {\rm M}_\odot \leq 10^9$, $10^9 < M_* /
{\rm M}_\odot \leq 10^{10}$, and galaxies above $M_* > 10^{10}$ M$_\odot$. The
lines show the mean density, stacked across all galaxies in the mass range, with
the final calculated mean the median of 1024 bootstraps (with replacement) of
the original selection. The bootstrapping is used to estimate the error on the
mean density, with the 16--84 percentile range shown as dashed lines. We also
show the dark matter profiles of the same galaxies, stacked in the same way, as
lighter lines in the background.

Fig. \ref{fig:stellarmassprofile} shows that backsplash galaxies have a significantly
lower central stellar density than their infalling counterparts of the same mass.
They also show a much more centrally concentrated density profile, in contrast to
the quenched infalling galaxies which have a strong cusp in the central stellar
density. These differences in stellar profiles are despite the dark matter profiles
of the two classes of galaxies being extremely similar.

As in each cut the masses of the galaxies are constrained to be similar, we can see
that although the sizes of the quenched infalling galaxies and backsplash galaxies
are similar, these originate through very different distributions. The quenched galaxies
have a much steeper profile that drops off rapidly, whereas the backsplash galaxies
have a shallower profile that catches up and overtakes the quenched profile at large
radii. 

The flat central profiles of the backsplash galaxies are consistent with those seen in 
\citet{Carleton2019}, where the central stellar core is tidally heated through
close interactions with the cluster halo as the galaxies orbit. Tidal heating injects
energy into particles within the galaxy as it falls into a cluster on a (crucially)
eccentric orbit, with the shear forces on the bound particles changing rapidly with
time. Tidal heating of backsplash galaxies is further consistent with the increased
sphericity seen in the shapes of backsplash galaxies in prior works \citep{Knebe2020}.

Such tidal interactions have been suggested to be the origins of ultra-diffuse
galaxies. A number of authors \citep[e.g.][]{Benavides2021, Trujillo2021} have
recently suggested that tidal heating and stripping processes could play a role
in forming such UDGs. The suppression of the central stellar density of the
backsplash galaxies identified in this study certainly make them UDG candidates,
and further suggest that a large number of UDGs near clusters may well be
backsplash galaxies. We caution, however, that the suppression of the central
stellar density occurs close to the softening scale of the TNG300 simulation,
making the close gravitational interactions less accurate than would be ideal
for studying such complex dynamics. These results suggest that follow-up studies
should focus on these low-mass backsplash galaxies, as they are potentially
the most distinct from their infalling counterparts, but due to the limited
resolution of TNG300 we are unable to make direct comparisons to nature.

\begin{figure}
    \centering
    \includegraphics{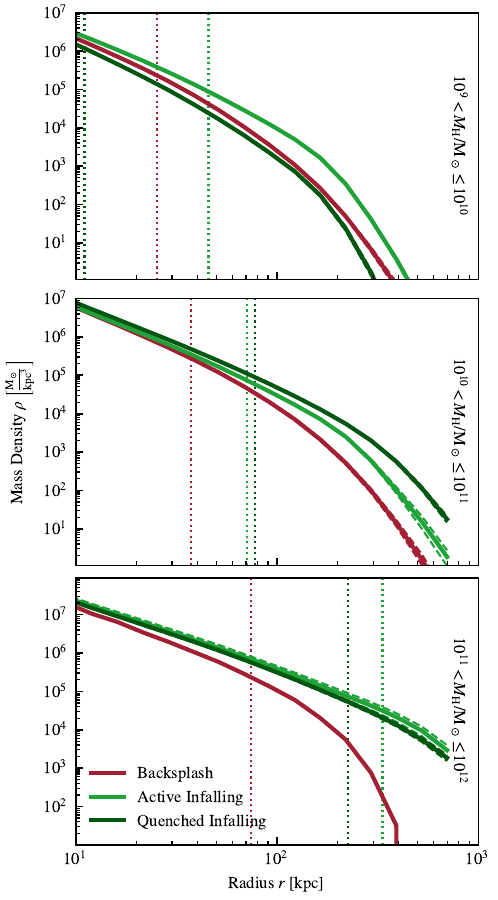}
    \caption{Mean total density profiles, split by final bound mass ($M_{\rm
    H}$; panels). Solid lines show the median, with the dashed lines showing the
    16--84 percentile range, of 1024 mean-stacked bootstraps (with replacement)
    of all individual profiles. The dotted lines show the median bound mass
    half-mass radius for galaxies in each mass bin of the corresponding colour.
    Note that the horizontal range is shifted significantly from the same figure
    for the stellar component (Fig.  \ref{fig:stellarmassprofile}). The inner
    regions are not shown as the profiles of all cuts are all consistent in dark
    matter and deviations are powered by differences in the stellar component.}
    \label{fig:total_mass_density_profile}
\end{figure}

In Fig. \ref{fig:total_mass_density_profile} we show the outer (total mass,
though in this region the mass density is dominated entirely by dark matter)
profiles of our haloes. Here, we split by halo mass, but only include galaxies
that are occupied by at least $M_* > 10^8$ M$_\odot$ for consistency. We see
that at masses $M_{\rm H} > 10^{10}$ M$_\odot$ there is significant disruption
of the backsplash subhaloes, leading to smaller dark matter half-mass sizes (as
can be seen in Fig. \ref{fig:stellarmassprofile}, the inner dark matter profiles
are consistent for backsplash and infalling galaxies). This suggests that there
has been significant tidal stripping of subhaloes as they have passed through
the cluster, with the strength of the stripping increasing with increasing
subhalo mass. This is consistent with the results from \citet{Jiang2016} who studied
the disruption of subhaloes in the Bolshoi simulation, and found that the
tidal mass loss rate, $\mathrm{d}M_{\rm H} / \mathrm{d}t \propto M_{\rm H}^{1.07}$,
shows superlinear growth with subhalo mass.

\citet{vandenBosch2017}, \citet{vandenBosch2018a}, and \citet{vandenBosch2018b}
all discuss the potential origins of forces leading to tidal stripping, and
whether the disruption of substructure is numerical or physical in nature.
Notably, they find that disruption of subhaloes is converged over many softening
scales and resolutions on short timescales (comparable to one or two cluster
crossing times, see Fig. 10 in \citet{vandenBosch2018b}), but subhaloes continue
to lose mass due to numerical effects on long timescales.
\citet{vandenBosch2018b} focused only on subhaloes on small ($R / R_{\rm vir}
\approx 0.15$) circular orbits, however, meaning that in realistic scenarios
(like our backsplash subhaloes), the weaker tidal fields may have a less
significant effect. As our backsplash galaxies are typically on their way to
their first periapsis (i.e. they have only been influenced by the cluster for a
relatively short amount of time), we suggest that the tidal disruption that our
subhaloes have experienced is likely to be physical at large radii.

In summary, we have found that the stellar sizes of backsplash galaxies are
larger than the general field population, especially at low masses, whilst the
bound mass (or dark matter mass) sizes of these galaxies are smaller than the
field population, for a given subhalo or stellar mass. We have discussed how
these differing sizes are likely due to tidal interactions with the host
cluster: as galaxies fall in to the cluster, their cores ($r < 5$ kpc) are
tidally heated (leading to a flatter inner stellar density profile, and larger
stellar size, see Fig. \ref{fig:stellarmassprofile}), whilst their extremities
($r > 100$ kpc) are tidally stripped (see Fig.
\ref{fig:total_mass_density_profile}). This leaves us with galaxies that are
dark matter dense, but gas and star poor, making the backsplashing of galaxies a
potential UDG formatiom mechanism, as suggested by prior authors.

% {\color{red}
% Potential further discussion points:
% \begin{itemize}

% \item Better describe what tidal heating is and how it works?

% \item Consider trying f(stellar mass) as a function of radius, this will help
%       considerably with any normalisation issues?

% \item Maybe we should look at the eccentricity of the orbits here?
% \item Note that the density of the DM in the center is the same as stars in low
%       mass galaxies?
% \end{itemize}
% }
\section{Black Hole Occupation}
\label{sec:blackholes}

\begin{figure}
    \centering
    \includegraphics{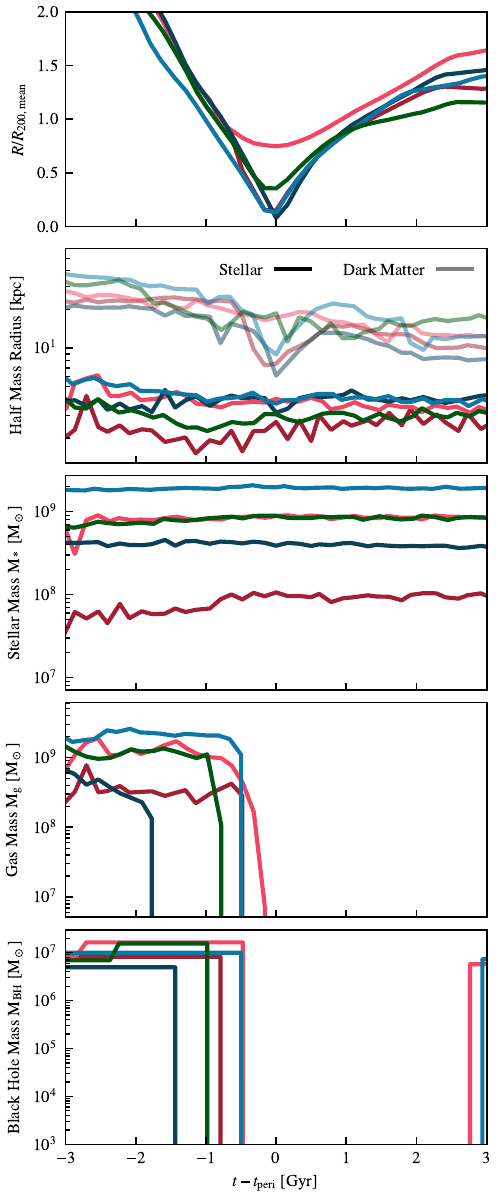}
    \caption{Tracks of cluster galaxies, rescaled to ensure that their
    pericentres overlap in time, all falling into the same ($M_{\rm 200, mean} =
    2.13 \times 10^{15}$ M$_\odot$ cluster). Each line (different colours) shows
    a different infalling galaxy). From top to bottom, the panels show the
    orbital radius of the galaxies, their stellar and dark matter half-mass
    radii, their bound stellar mass within twice the (stellar) half-mass radius,
    their bound gas mass (note when lines leave the bottom of the plot they have
    a value of zero) in the same aperture, and the total bound black hole mass,
    as a function of time.}
    \label{fig:masstracing}
\end{figure}

To better understand both the radial profile differences explored in \S
\ref{sec:sizes}, and the importance of the black hole mass in classifying
backsplash galaxies, we now consider the evolution of individual backsplash
galaxies.

In Fig. \ref{fig:masstracing} we show the properties of backsplash galaxies falling
into the most massive cluster in the volume ($M_{\rm 200, mean} = 2.13 \times
10^{15}$ M$_\odot$ cluster). These five galaxies are selected randomly through
the mass range, and are chosen to have one pericentre before residing outside of
the cluster at $z=0$. 

In the top panel of Fig. \ref{fig:masstracing} we show the radial tracks of the
galaxies as they fall into, and rebound from, the cluster. To ensure that we can
consistently compare the galaxies, we re-scale the time axis by the time at
which each galaxy reaches the pericentre of their orbit $t_{\rm peri}$. This
time is computed as the output time of the snapshot at which the radius
$R(t)/R_{\rm 200, mean}(t)$ is at a minimum. We see that there are some galaxies
present that only have a glancing orbit (pink curve) and some that have
pericentres very close to the centre of the cluster, such as the dark and light
blue curves.

The second panel of Fig. \ref{fig:masstracing} shows the half-mass sizes for
stars (dark) and dark matter (light). We see that as low mass galaxies (red and
green) pass pericentre, their stellar sizes are inflated, and all galaxies have
their dark matter half-mass sizes reduced during their time in the cluster.
This further strengthens our hypothesis from \S \ref{sec:sizes} that tidal
interactions with the cluster control these two variables.

The third panel in Fig. \ref{fig:masstracing} shows the stellar masses of the
galaxies over time, here showing the stellar mass within twice the half-mass
radius. That there is little evolution in this quantity over time, even at
pericentre passage, suggests that rather than explicitly losing mass to produce
a more cored stellar profile (Fig. \ref{fig:stellarmassprofile}) the stellar
mass in the centres of the galaxies is simply rearranged, strengthening our
hypothesis that this rearrangement occurs due to tidal heating.

In the fourth panel, we show the gas mass within twice the stellar half-mass
radius for each of the galaxies. We see that by the time that the galaxies have
reached the pericentre of their orbit they have all lost all of their gas
through ram pressure stripping from the cluster. This even occurs in the
galaxy with a relatively shallow orbit (pink curve, $R_{\rm peri} \approx 0.8
R_{\rm 200, mean}$), suggesting that ram pressure stripping is very efficient
in this massive cluster. In idealised simulations of galaxies in `wind tunnels',
\citet{Tonnesen2009} found that galaxies are stripped of over half of their gas
mass in under 0.5 Gyr upon entry into a cluster-like environment.
\citet{Simpson2018} studied backsplash galaxies (dwarfs orbiting local groups)
in simulations, and found that even in these shallow potentials satellites can
be completely stripped of gas.

Finally, in the bottom panel, we show the black hole mass of the galaxies as a
function of their orbital time. On a similar timescale to their gas loss, the
cluster galaxies appear to lose their black holes. This is unexpected, as the
black holes act as collisionless particles and typically reside in the centres
of the cluster galaxies which we have seen in the second panel ($M_*$) are
relatively protected aside from tidal heating. Further, black hole particles are
`pinned' to the centres of their galaxies by a `repositioning' scheme that
attempts to insulate them from spurious numerical heating \citep[see][Section
2.2]{Weinberger2017}. In addition, we see that once the galaxies have moved away
from the cluster (at $t - t_{\rm peri} > 2.5$ Gyr) they appear to re-grow black
holes, but at larger masses than the black hole seed mass in the simulation
\citep[$1.1 \times 10^6$ M$_\odot$, see][]{Pillepich2018}.

\begin{figure}
    \centering
    \includegraphics{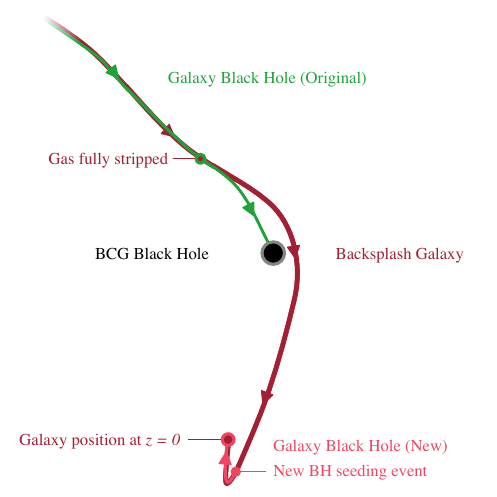}
    \caption{A schematic showing the stripping of black holes from their host
    backsplash (or cluster) galaxies. The dark red line shows the path of the
    backsplash galaxy, in the reference frame of the black hole at the centre
    of the central brightest cluster galaxy (BCG, black circle). The green line
    shows the original black hole associated with the backsplash galaxy, which is
    removed by the repositioning scheme once the galaxy is stripped of all of
    its gas. Once back outside the cluster, the backsplash galaxy is re-seeded
    with a new black hole (pink line), before residing outside the cluster at
    $z=0$.}
    \label{fig:schematic}
\end{figure}

It appears in this case that the repositioning scheme, initially intended to
ensure that the positions of black holes remained stable, is the downfall of
black holes living within cluster galaxies. The black hole repositioning scheme
is designed to ensure that the particle is repositioned onto the particle (of
the nearest 1000) with the lowest gravitational potential\footnote{We note here
that this is different from what is reported in \citet{Pillepich2018}, who
report that only gas particles are used. After a thorough investigation of the
code with the original developers, we found that all particle types are
included.}. When repositioned, the black hole particle is given the same
velocity relative to the volume as the particle it is repositioned onto. If the
galaxy is a central, e.g. the brightest cluster galaxy (BCG), then such a scheme
keeps the black hole pinned to the potential minimum. In a cluster galaxy that
is undergoing stripping, however, this leads to the black hole being
repositioned onto either a particle with a high relative velocity to the halo,
or a member bound to, and phase mixed with, the cluster, making it leave the
substructure.  This eventually leads to the rapid movement of the black hole to
the centre of the cluster, merging with the black hole at the centre of the BCG.
Fig. \ref{fig:schematic} shows a schematic, based upon the path of the light
blue trace in Fig. \ref{fig:masstracing}, of how a black hole is lost and
re-seeded in a backsplash galaxy.

\begin{figure}
    \centering
    \includegraphics{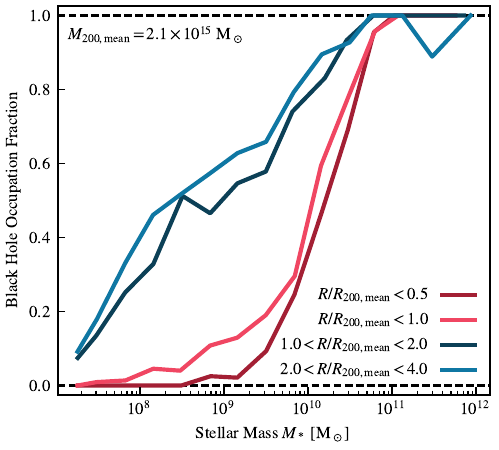}
    \caption{Shows the occupation fraction of black holes, computed as the
    number of galaxies in this bin that have a black hole divided by the total
    number of galaxies, for 16 equally log-spaced bins between the figure
    limits.  Each line shows a different aperture around the cluster within
    which the occupation fraction is computed. The occupation fractions here are
    shown for the most massive cluster in the volume, and the same cluster shown
    in Fig. \ref{fig:masstracing}, with $M_{\rm 200, mean} = 2.13 \times
    10^{15}$ M$_\odot$, though these trends occur across the mass range.}
    \label{fig:bhocc}
\end{figure}

Fig. \ref{fig:bhocc} shows the occupation fraction of black holes as a function
of their stellar mass, in various apertures. The blue lines show apertures
generally containing field and infalling galaxies, with the red lines showing
the cluster galaxy component. Notably, the occupation fraction of galaxies
within the cluster is significantly lower, especially at lower masses (those
expected to survive post-apocentre to become backsplash galaxies). Below $M_* <
10^{10}$ M$_\odot$, galaxies within the core of the cluster ($R/R_{\rm 200,
mean} < 0.5$) have occupation fractions that are roughly zero, suggesting all of
these (there are 1243 galaxies with $10^{7} < M_* / {\rm M}_\odot < 10^{10}$)
galaxies have lost their black holes. The higher mass galaxies are those less
likely to become backsplashers, as they generally rapidly merge with the central
due to dynamical friction \citep{Bakels2021}, and as such are less likely
to have already had their pericentre and lost their black hole.

Recent works, such as \citet{Bahe2021}, have noted that there appears to be a
lower than expected black hole occupation fraction for cluster galaxies (see
their Appendix A). Notably, the model discussed in \citet{Bahe2021} also
repositions only on gas particles, as typically the neighbour-finding structures
(e.g. trees) within N-body codes are only set up to include gas particles for
optimisation reasons. Our results, along with those in \citet{Bahe2021}, suggest
that the black hole occupation fraction of cluster galaxies is severely
underestimated within cosmological simulations that employ such repositioning
schemes.

Such issues with the repositioning of black holes are likely to persist amongst
many of the state-of-the-art large-volume simulations, including the entire
Illustris and IllustrisTNG suite, as well as the EAGLE suite and derivatives
\citep{Vogelsberger2014, Schaye2015}. Simulations that do not employ
repositioning, such as the Horizon-AGN suite \citep{Dubois2012}, or Romulus
\citep{Tremmel2017}, will not suffer from this artificial stripping of black
holes from cluster galaxies. Prior work from \citet{Ragone-Figueroa2013}
employed repositioning on all particle types, though unlike TNG has black hole
growth dominated by accretion at all masses \citep{Bassini2019}, with TNG black
holes showing merger-dominated growth at $M_{\rm BH} > 10^8$ M$_\odot$
\citep{Weinberger2018}. It is possible that these authors, like the EAGLE
simulations, included a maximal velocity difference between the black hole and
repositioning candidates, but a full investigation of the implementation of
black hole recentreing strategies is out of the scope of this work.
Additionally, modifying the neighbour finding structures used for repositioning
to employ stellar particles for recentreing purposes would insulate such schemes
from these effects. The SIMBA simulation suite, that employs a loop over nearby
stellar particles to compute accretion rates, already uses a recentreing
prescription that includes the use of stellar particles, and likely does not
suffer from these black hole stripping issues \citep{Angles-Alcazar2017}.

\citet{Haidar2022} investigated the occupation fraction of black holes in a
number of cosmological simulations, and found that the SIMBA suite, that
repositions on stars as well as gas particles, have occupation fractions of
close to 100\% even down to low masses. They find that EAGLE and IllustrisTNG
actually reduce their global occupation fractions of low-mass galaxies over
time, due to the increasing fraction of galaxies resident in clusters. We have
now demonstrated the specific numerical pathway through which these galaxies are
stripped of their black holes.

There are a number of potential pitfalls for galaxy formation simulations with
black hole stripping. Primarily, aside from the black hole occupation fractions
being clearly incorrect (the artificial seeding procedure makes these numbers
uncertain anyway), the contribution to the cluster energy budget from the AGN
resident in cluster galaxies will be underestimated.  Additionally, the merger
rates of black holes, particularly those in the centres of clusters, will be
overestimated, as will their masses.  The impacts of these outcomes are not yet
known, but as black hole seeding and AGN models are explicitly calibrated
against observational data \citep[e.g.][]{Crain2015,Pillepich2018} and as such
these potential pitfalls may already be calibrated into the choice of free
parameters.
\section{Conclusions}
\label{sec:conclusions}

Backsplash galaxies are galaxies that have previously had a pericentre within
the bounds of a galaxy cluster, but have since migrated outside of most typical
definitions of the cluster boundary. Their intrinsic properites are not well
understood, as simulating such galaxies is numerically challenging due to the
high inherent dynamic range of masses and spatial scales present in galaxy clusters.

In this paper we have explored the properties of backsplash galaxies in the
TNG300 simulation by tracing the orbits of galaxies with stellar mass $M_* >
10^8$ M$_\odot$ within $1 < R/ R_{\rm 200, mean}<10$ of our 1302 isolated clusters with
$M_{\rm 200, mean} > 10^{13}$ M$_\odot$. By classifying galaxies into `backsplash'
and `infalling' categories based upon whether the galaxy has previously been within
$R_{\rm 200, mean}$ of the cluster re, we were able to find that:
\begin{itemize}
    \item The abundance of backsplash galaxies grows roughly linearly
          as a function of galaxy cluster mass, with $M_{\rm 200, mean} \approx
          10^{14}$ M$_\odot$ clusters on average hosting around 5 backsplashers,
          and clusters with $M_{\rm 200, mean} \approx 10^{15}$ M$_\odot$
          hosting around 50 on average. Backsplash galaxies are significantly
          more common (relative to infalling galaxies) at smaller radii, with
          equipartition reached between the two populations at $R = 1.24R_{\rm
          200, mean}$. These two results suggest that observations targeting
          backsplash galaxies should focus on the intermediate regions ($1 <
          R/R_{\rm 200, mean} < 1.5$) of the highest mass clusters $M_{\rm 200,
          mean} > 10^{14.5}$ M$_\odot$, to reduce contamination and ensure
          adequate sampling.
    \item We verify that backsplash galaxies and infalling galaxies are kinematically
          separated, with the separation increasing as a function of host
          cluster mass. Both backsplash and infalling galaxies have a roughly
          gaussian radial velocity ($v_{\rm r}$) distribution, with infalling
          galaxies having a mean velocity $\bar{v}_{\rm r} < 0$ dependent on
          cluster mass ($\bar{v}_{\rm r} = -1100$ km s$^{-1}$ for $M_{\rm 200, mean}
          \approx 10^{15}$ M$_\odot$). Backsplash galaxies have a typical mean
          radial velocity of $\bar{v}_{\rm r} \approx 0$, with the width of the
          distribution dependent on cluster mass. These findings confirm that
          the use of phase-space information is an excellent way to classify
          backsplash galaxies, if it is available \citep[e.g. as
          in][]{Farid2022}.
    \item By considering typical galaxy scaling relations, we have shown that
          backsplash galaxies in TNG300 are typically black hole-poor, have
          higher mass-to-light ratios, larger stellar sizes, redder colours,
          and lower gas fractions than their infalling counterparts of a similar
          stellar mass. These characteristics are generally consistent with
          galaxies that have been ram pressure stripped, quenched, and
          tidally heated, processes thought to be common in galaxy clusters, and
          that have previously been shown to impact the evolution of 
          backsplash galaxies \citep[e.g.][]{Simpson2018}.
    \item To investigate which intrinsic properties of galaxies (without considering
          phase-space information) are key to separating the infalling and
          backsplashing population we trained a decision tree classifier on all
          available (reasonable) properties from the TNG SubFind catalogues of
          our two populations. The classifier identified the most important
          features in separating the two populations of galaxies as the galaxy
          gas mass, distance from the cluster re, galaxy and halo size,
          magnitude, and black hole mass. Our decision tree classifier was able
          to classify backsplash and infall galaxies with high accuracy,
          returning a 20\% cross-validation score of 89\%. When considering
          backsplash galaxies alone, the classifier had a low (7\%)
          false-positive rate, and a relatively low false-negative rate (22\%).
          These levels of accuracy are comparable with other hard classifiers that 
          do employ phase-space information, such as those used recently by
          \citet{Farid2022}. Despite the high accuracy of the classifier,
          however, it would be impractical to use for observations, and as
          such we simply use the decision tree as a way to reduce the
          extreme dimensionality of our feature space, finding interesting
          galaxy properties to focus follow-up analysis on.
    \item Through the analyis of galaxy sizes and profiles, motivated by
          the decision tree, we showed that backsplash galaxies have typically
          larger stellar sizes at low mass, and smaller dark matter sizes at
          high mass, than their infalling counterparts, even when accounting for
          the size growth of quenched galaxies due to numerical effects
          \citep{Genel2018}. We discussed how, despite numerical concerns with
          (relatively) low resolution simulations like TNG300, our results
          still suggest that the central regions of backsplash galaxies are
          tidally heated, leading to flatter stellar inner density profiles.
          They also suggest that backsplash galaxies are subject to tidal
          stripping, with the outskirts of backsplash galaxies relatively devoid
          of dark matter relative to their infalling counterparts. We discussed
          how these processes contribute to the potential formation of UDGs near
          clusters, where UDGs are likely to be backsplash galaxies.
    \item We investigated the low black hole occupation fraction of backsplash
          galaxies by tracking the detailed properties of individual galaxies
          back over time. We showed that backsplash galaxies typically lose all
          of their gas mass before reaching the pericentre of their orbit around
          the cluster, even when their orbit is shallow. We also showed how the
          black hole repositioning (or re-reing) prescription, which moves
          black hole particles within galaxies based on the local potential
          sampled by gas particles, leads to the extraction of black holes from
          infalling galaxies as they near the pericentre of their orbit. Such
          repositioning schemes, required to insulate black holes from spurious
          heating out of central galaxies and to maintain the efficiency of AGN
          feedback, hence come at the cost of underestimating the occupation
          fraction of black holes in cluster galaxies and hence
          their feedback contribution within clusters.
\end{itemize}
The large sample size of isolated massive haloes ($M_{\rm 200, mean} > 10^{13}$
M$_\odot$) in TNG300 has enabled us to study a representative sample (over
5000) of backsplash galaxies. We have found that the properties of these
galaxies are crucially dependent on interactions with the host cluster that are
still not well understood (or well modelled) in simulations, namely: tidal
heating, tidal stripping, and ram pressure stripping. Thus, we find that, in
general, backsplash galaxies are an excellent marker in the sand for comparison
to observations. In future work, higher resolutions and ever larger sample
sizes, along with expanded observational samples from the upcoming DESI, Nancy
Grace Roman Space Telescope, and Vera C. Rubin Observatory will provide a
fruitful ground for exploration of galaxy formation models through the study of
backsplashers.

\section*{Acknowledgements}

The authors thank Aaron Ludlow for conversations that significantly improved
the discussion on galaxy sizes.
The authors thank Yannick Bah\'e, R\"udiger Pakmor, Romeel Dav\'e, and
Rahul Kannan for conversations that significantly improved the discussion black
hole loss for cluster galaxies.

Some of the computations in this paper were run on the FASRC Cannon cluster
supported by the FAS Division of Science Research Computing Group at Harvard
University. Some of the computations were performed on the Engaging cluster
supported by the Massachusetts Institute of Technology. MV acknowledges support
through NASA ATP 19-ATP19-0019, 19-ATP19-0020, 19-ATP19-0167, and NSF grants
AST-1814053, AST-1814259, AST-1909831, AST-2007355 and AST-2107724.
AS acknowledges support for Program number \textit{HST}-HF2-51421.001-A provided
by NASA through a grant from the Space Telescope Science Institute, which is
operated by the Association of Universities for Research in Astronomy,
incorporated, under NASA contract NAS5-26555.

Software citations:
\begin{itemize}
	\item {\textsc{AREPO}}: \citet{Springel2011, Weinberger2020}
	\item {\textsc{Python}}: \citet{vanRossum1995}
	\item {\textsc{Matplotlib}}: \citet{Hunter2007}
	\item {\textsc{SciPy}}: \citet{Virtanen2020}
	\item {\textsc{Scikit-Learn}}: \citet{scikit-learn}
	\item {\textsc{NumPy}}: \citet{Harris2020}
	\item {\textsc{SwiftSimIO}}: \citet{Borrow2020, Borrow2021}
\end{itemize}

%%%%%%%%%%%%%%%%%%%%%%%%%%%%%%%%%%%%%%%%%%%%%%%%%%
\section*{Data Availability}

The data underlying this article are available in Zenodo at
{\url{https://dx.doi.org/10.5281/zenodo.6564713}}. All data in this article were
reduced from the publicly available TNG300-1 simulation data available at
{\url{https://tng-project.org}} \citep{Nelson2019}.

%%%%%%%%%%%%%%%%%%%% REFERENCES %%%%%%%%%%%%%%%%%%

% The best way to enter references is to use BibTeX:

\bibliographystyle{mnras}
\bibliography{bibliography} % if your bibtex file is called example.bib

% Alternatively you could enter them by hand, like this:
% This method is tedious and prone to error if you have lots of references
%\begin{thebibliography}{99}
%\bibitem[\protect\citeauthoryear{Author}{2012}]{Author2012}
%Author A.~N., 2013, Journal of Improbable Astronomy, 1, 1
%\bibitem[\protect\citeauthoryear{Others}{2013}]{Others2013}
%Others S., 2012, Journal of Interesting Stuff, 17, 198
%\end{thebibliography}

%%%%%%%%%%%%%%%%%%%%%%%%%%%%%%%%%%%%%%%%%%%%%%%%%%

%%%%%%%%%%%%%%%%% APPENDICES %%%%%%%%%%%%%%%%%%%%%

\appendix
\section{Feature Importance at Higher Masses}

\begin{figure}
    \centering
    \includegraphics{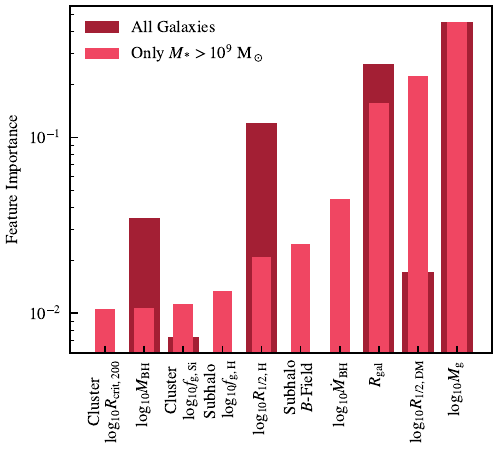}
    \caption{Same as Fig. \ref{fig:feature_importance}, but only including
    galaxies with $M_* > 10^9$ M$_\odot$ in the training set. We see very
    similar important features, notably the gas mass, distance from cluster
    centre, and dark matter half-mass size. In the background, in the original
    red, we show the top 10 feature importances as displayed in the original
    figure.}
    \label{fig:feature_importance_1e9}
\end{figure}

In Fig. \ref{fig:feature_importance_1e9}, we show the results of our machine
learning pipeline from Section \ref{sec:classifier}, applied to galaxies with a
stellar mass $M_* > 10^9$ M$_\odot$ only, to test how our results change when
considering what would be typically considered `well resolved' galaxies only.
The important features are nearly identical to those when we train the pipeline
on all substructures as in Fig. \ref{fig:feature_importance}, with the only
major missing important feature the stellar half-mass sizes. This is expected,
as demonstrated in Fig. \ref{fig:stellarmassprofile}, the differences in stellar
profiles were mainly powered by the lowest-mass galaxies and were only
marginally resolved in the TNG-300 simulation.

The black hole mass feature is replaced with the black hole accretion rate, but
again this is simply used for occupation checking. The final, somewhat
important, feature that is new here is the Subhalo $B-$field. This feature is
somewhat degenerate with the gas mass, with backsplash haloes that retain gas
having relatively low magnetic field strengths due to their low gas fraction.

As the same qualitative trends remain from our main results, and there are
potentially interesting features shown in the $10^8 < M_* / {\rm M}_\odot < 10^9$
mass range, in the main text we retain the larger sample.

%%%%%%%%%%%%%%%%%%%%%%%%%%%%%%%%%%%%%%%%%%%%%%%%%%

% Don't change these lines
\bsp	% typesetting comment
\label{lastpage}
\end{document}